\documentclass[journal]{IEEEtran}
\usepackage{amsmath}
\usepackage{cuted}
\usepackage{xcolor}
\usepackage{amsmath,amssymb}
\usepackage{amsthm}
\newtheorem{theorem}{Theorem}
\newtheorem{remark}{Remark}
\usepackage[compact]{titlesec}   
\usepackage{upgreek}
\usepackage{graphicx}
\usepackage{booktabs}
\usepackage{setspace}
\usepackage{hyperref} % <--- Load hyperref near the bottom
\hypersetup{
    colorlinks=true,
    linkcolor=blue,
    citecolor=blue,
    urlcolor=blue
}
\usepackage{overpic}
\usepackage{amsthm}
\begin{document}

\title{\huge Cross-product Modelling in Continuous Intraday Electricity Markets via Hawkes Processes}

\author{Ayoub Jhabli$^{\ast}$$\dag$\thanks{$^\ast$Corresponding author. Email: ayoub.jhabli@\{wur.nl, cern.ch\}}, Tarek AlSkaif$^{\dag}$, Kwabena E. Bennin$^{\dag}$, Bedir Tekinerdogan$^{\dag}$, and Joost M.E. Pennings$^{\S\P}$\\
{\normalsize\itshape $\dag$Information Technology Group, Wageningen University, The Netherlands\\
$\S$Marketing and Consumer Behaviour, Wageningen University, The Netherlands\\
$\P$School of Business and Economics, Maastricht University, The Netherlands} 
 
    \thanks{
        This research is fully funded by the Starter Grant awarded to Dr. Tarek Alskaif and Dr. Kwabena Bennin at the Information Technology Group, Wageningen University. It is partly supported by Project HighLO a collaboration between Wageningen University \& Research (WUR) and the European Organization for Nuclear Research (CERN), which aims to identify malicious trading behavior such as market manipulation, to better protect markets.}}    
%\author{\IEEEauthorblockN{Ayoub Jhabli\IEEEauthorrefmark{1}\IEEEauthorrefmark{4}, Tarek Alskaif\IEEEauthorrefmark{1}, Kwabena E. Bennin \IEEEauthorrefmark{1}, \\ Bedir Tekinerdogan\IEEEauthorrefmark{1}, Axel Naumann \IEEEauthorrefmark{3} \& Joost M.E. Pennings \IEEEauthorrefmark{2} \\ \IEEEauthorrefmark{1}Information Technology Group, Wageningen University, The Netherlands,\\
%\IEEEauthorrefmark{2}Marketing and Consumer Behaviour, Wageningen University, The Netherlands\\
%\IEEEauthorrefmark{3}European Organization for Nuclear Research (CERN), Geneva, Switzerland\\
%\IEEEauthorrefmark{4}Corresponding author: ayoub.jhabli@wur.nl
%}} 
%\author{\IEEEauthorblockN{Ayoub Jhabli\IEEEauthorrefmark{1}\IEEEauthorrefmark{4}, Tarek Alskaif\IEEEauthorrefmark{1}, Kwabena E. Bennin \IEEEauthorrefmark{1}, \\ Bedir Tekinerdogan\IEEEauthorrefmark{1}, Axel Naumann \IEEEauthorrefmark{3} \& Joost M.E. Pennings \IEEEauthorrefmark{2} \\ \IEEEauthorrefmark{1}Information Technology Group, Wageningen University, The Netherlands,\\
%\IEEEauthorrefmark{2}Marketing and Consumer Behaviour, Wageningen University, The Netherlands\\
%\IEEEauthorrefmark{3}European Organization for Nuclear Research (CERN), Geneva, Switzerland\\
%\IEEEauthorrefmark{4}Corresponding author: ayoub.jhabli@wur.nl
%}}
\maketitle
\begin{abstract}
As renewable energy integration accelerates, intraday continuous (IDC) electricity markets have become essential for short-term balancing. However, the simultaneous trading of overlapping contracts with varying granularities creates complex, hidden liquidity dependencies that are critical to market stability but remain poorly understood. This paper proposes a multivariate State-Dependent Hawkes Process (SDHP) to model cross-product liquidity dynamics between Quarter-Hourly and Hourly IDC contracts. To determine how cross-product effects vary under different market conditions, we classify the trading sessions into two distinct regimes: 'Normal' and 'Stressed', based on a composite metric of the bid-ask spread, mid-price volatility and queue imbalance. The model is calibrated on the first month of each quarter of 2024 using level 3 limit order book (LOB) data from the Dutch EPEX SPOT IDC market. We find that order flow is overwhelmingly self-driven, with roughly 95\% of orders arriving in reaction to earlier orders in both regimes. Cross-excitation runs mainly from the Quarter-Hourly to the Hourly contract, but this leadership weakens under stress: the ratio of cumulative cross-excitation falls from 2.8 to 1.5, so the two contracts influence each other almost equally. Stress also makes the Quarter-Hourly to Hourly channel much faster, cutting its mean lag from 6.05 to 0.93 seconds. The strong seasonal variation found in calm periods largely disappears under stress, where all four months converge to the same structure.
\end{abstract}


%\item 
%\item 
%\item 
% NOTE: No highlights were provided in the original manuscript; none have been added here.

% Keywords

\section{Introduction}
\label{sec:intro}

With rising shares of wind and solar generation, the intraday continuous (IDC) market has become the main venue for correcting imbalances between the day-ahead auction and real-time balancing \cite{eurelectric2026shortterm}. Participants use it to adjust positions after the day-ahead result is known, typically in response to updated wind and solar PV power forecasts \cite{visser2024probabilistic}, plant outages or revised demand \cite{le2019integrated}. Unlike the day-ahead auction, which clears all orders at a single price, IDC markets match orders one by one as they arrive. A trade occurs whenever a buy and a sell order meet, so a single delivery period generates a sequence of prices rather than one, which exchanges summarise through volume-weighted indices such as EPEX's IDFull, ID3 and ID1 \cite{SERAFIN2022106125, NARAJEWSKI2020100107, Grindel2024, NARAJEWSKI2020115801}.

At European level, continuous intraday trading is coordinated through the Single Intraday Coupling (SIDC) project, implemented on the M7 Cross-Border Intraday (XBID) platform \cite{Vogeler2025, entsoe2026sidc}. XBID connects the Nominated Electricity Market Operators through a central Shared Order Book (SOB), where market participants across Europe are able to trade electricity continuously, subject to available transmission and interconnection capacity. Contracts are classified as global, and therefore eligible for cross-border matching in the SOB, or local, and matched only on the domestic platform within their respective zone. Matching in the SOB is constrained by available cross-zonal transmission capacity, which is tracked by the Capacity Management Module (CMM). The SIDC system has grown quickly: quarterly executed trades rose from 3.50 million in Q3 2018 to 73.26 million in Q2 2025 \cite{entsoe_sidc}.

EPEX SPOT, one of the largest European electricity exchanges, lists contracts of several granularities for intraday continuous trading: Hourly, Half-hourly and Quarter-hourly individual contracts going 24/7 all year round \cite{epex_brochure}. Trading takes place through a limit order book (LOB), which makes the market structurally similar to an electronic financial market, though with important differences arising from the physical delivery \cite{8469829}.

Liquidity determines how cheap trading can be. In a liquid book a participant can correct an imbalance quickly at a price close to the prevailing mid; in an illiquid one the same correction moves the price against them or cannot be executed at all. Liquidity in power markets has therefore been studied through the traded volume, bid-ask spread, market depth and resiliency \cite{hagemann2013}. Empirical work on the German market shows that bid-ask spreads widen with risk and narrow with trading activity and competition close to delivery \cite{8eb9b2db-1a97-3f35-a1c8-dfe7e38130c7}. Liquidity is also uneven across granularities: Quarter-hourly products are consistently thinner than Hourly products, which is why much of the forecasting and execution literature restricts attention to the Hourly contracts.

The observation that contract dynamics change as delivery approaches is not specific to electricity. Samuelson \cite{samuelson1965} argued that the volatility of a futures contract rises as it nears maturity, because information about the terminal value arrives faster and matters more. In electricity this mechanism is present as well. Weather forecasts sharpen as delivery approaches, imbalances become known, and participants must close their positions before gate closure \cite{Equilibrium2022}. Empirical studies confirm the pattern in IDC power markets: volatility, trade frequency and order book activity all increase towards gate closure \cite{Chatziandreou08012026,decharte2023}. The intensity of orders can therefore not be treated as constant through a session.

Point processes provide a natural way to model the intensity of orders, because they describe events in continuous time rather than on a fixed grid. Hawkes process, introduced to model earthquake aftershocks \cite{496a8ed0-5e76-38ab-8d67-eaa3d8cd00d7}, is the standard choice when events cluster. Its defining feature is self-excitation: each event excites further events for a period afterwards, with the increase decaying over time. The framework has been extended in several directions. Multivariate specifications treat several event types jointly and estimate the full matrix of self- and cross-excitation \cite{BOWSHER2007876,Embrechts_Liniger_Lin_2011}. Marked Hawkes processes attach a variable to each event, such as a trade size or a jump magnitude, allowing the strength of excitation to depend on it \cite{LEE2017174}. Regime-switching and Markov-modulated specifications let the parameters vary with a latent state, capturing alternation between quiet and bursty periods \cite{wu2021markovmodulatedhawkesprocessessporadic}. State-dependent Hawkes processes (SDHP) instead couple the process to an observable state, so that excitation depends on the current condition of the market and the state itself evolves with the events \cite{Morariu-Patrichi04032022}.

% In the multivariate case the process also allows mutual excitation, so an event of one type can raise the intensity of another type (for example, an order in the hourly contract can excite subsequent order arrivals in a quarter-hourly contract). This makes it possible to measure not only how strongly a market reacts to itself but also how activity is transmitted between distinct event types or as in finance distinct instruments \cite{Hawkes01022018}.

In financial markets these models have been applied extensively to LOBs \cite{Chen03052022}. They have been used to estimate transaction-level volatility and to explain volatility clustering \cite{BOWSHER2007876}, to model the joint arrival of market orders, limit orders and cancellations \cite{repec:zbw:ifweej:201222}, to specify multitype marked Hawkes processes nonparametrically with state-dependent baselines driven by order book imbalance \cite{Kirchner03052022}, and to quantify how much activity is generated endogenously by the market itself rather than by external factors \cite{Filimonov_2012,Hardiman_2013}. That last strand is directly relevant here: branching-ratio estimates close to one imply that most observed activity is a reaction to previous activity, a property associated with long memory in order flow and with fragility, since shocks are damped only weakly \cite{10.1214/15-AAP1164}. A parallel literature models the order book through its queues rather than its event intensities, with state-dependent arrival rates that respond to the configuration of the book \cite{doi:10.1137/110856605,Huang02012015}.

Applications to electricity markets are more recent but growing. Hawkes models have been used to describe market order arrivals in the German IDC market, where self-excitation is found to be strong but short-lived and the estimated parameters vary systematically across delivery hours \cite{jrfm14040161}. They have been combined with transient price impact models to derive optimal execution strategies for participants trading large volumes, reproducing the intraday activity profile and the Samuelson effect \cite{chatziandreou2025optimalexecutionintradayenergy}. They have also been extended with exogenous covariates, showing that trading activity responds not only to its own past but to vRES forecast errors and to activated balancing volumes \cite{KRAMER2021105186}. Related point-process work models the transaction arrival process directly \cite{en12234518}, and Hawkes specifications have also been used to detect manipulative order flow \cite{fabre2025highfrequencymarketmanipulationdetection}. Alongside this, a separate literature studies intraday liquidity and order book structure without point processes, through spreads, depth and execution cost \cite{8eb9b2db-1a97-3f35-a1c8-dfe7e38130c7,KUPPELWIESER2021112299,Bergault03102025}.

The same contract can be traded in a liquid book with tight spreads for much of its session and in a thin, volatile book shortly afterwards, and conditions differ across delivery hours and seasons. If the way order flow propagates depends on these changing conditions, then a model with constant parameters estimates an average across regimes that may describe neither.

This is a problem for cross-product analysis in particular. Whether activity in the quarter-hourly book drives the hourly book, or the reverse, is a statement about the structure of transmission, and there is no reason to expect that structure to be the same when the book is deep and when it is nearly empty. A constant-parameter model cannot answer this question, because it imposes the same structure across different conditions. Allowing the baseline intensity and the excitation kernel in Hawkes models to depend on an observable measure of liquidity stress separates the two cases: exogenous order arrivals and the market's reaction to its own past are then estimated within each regime, and the difference between regimes becomes the object of study rather than an assumption. That conditioning on liquidity stress matters has already been shown for this market. In earlier work on Dutch intraday hourly products, regimes defined by a liquidity stress index built from the bid-ask spread, order book volume and mid-price volatility fitted the order flow considerably better than a single-regime benchmark, and order flow was found to be strongly self-exciting and near-critical \cite{jhabli2026regime}.

In the financial literature, the predominant methodology for measuring liquidity spillovers relies on the generalized vector autoregressive (VAR) framework and variance decomposition introduced by Diebold and Yilmaz \cite{DIEBOLD201257}. This macro-econometric approach has been widely adapted to quantify directional spillovers across various domains, such as assessing global risk in foreign exchange markets \cite{CHANG2022102105} and analyzing the simultaneous dry-up of liquidity between ETFs and their underlying constituent stocks \cite{SON2023723}. However, these variance decomposition methods operate on temporally aggregated, lower-frequency data, making them inadequate for modeling event-based data. Multivariate Hawkes processes have emerged as an alternative by capturing the mutually exciting nature of market shocks \cite{Schneider01022018, Pindza26}. Yet, these existing point-process applications in this context still fall short of capturing continuous market microstructure dynamics; they either integrate Hawkes intensities with aggregated DCC-GARCH frameworks for macro-level contagion \cite{Pindza26}, or they restrict the event space to extreme, threshold-triggered illiquidity drops rather than modeling the continuous arrival of high-frequency events \cite{Schneider01022018}. 

Three research gaps follow. First, cross-product spillovers in IDC electricity markets are largely understudied. Most empirical work studies a single granularity, usually the hourly contract. How much activity in one transmits to the other, in which direction, and how quickly, is not established.

Second, the effect of liquidity conditions on excitation across products is not understood. The endogeneity literature reports branching ratios close to one in financial markets, and \cite{jhabli2026regime} studies how stress reshapes excitation within a contract, how endogenous electricity markets are but it is unknown whether stress changes how much activity spills between different contracts, or in which direction.

Third, kernel choice is usually asserted rather than tested, so the number of timescales in the market's memory is imposed and not empirically established.

The SDHP model, the LSI-based regimes, the regime-dependent baseline and excitation, and the spectral-radius penalty follow \cite{jhabli2026regime}. This paper makes the following contributions.

\begin{itemize}
    \item \textit{Kernel comparison:} We replace the single-exponential kernel of \cite{jhabli2026regime} with three exponential-mixture kernels and a power-law kernel, and select the best performing one for economic interpretation.

    \item \textit{Estimation at scale:} We extend the analytical gradients of \cite{jhabli2026regime} to both kernel families, which makes it feasible to estimate the model separately for the first month of each quarter, covering 2,915 delivery contracts.

    \item \textit{Cross-product spillovers:} We measure spillovers between Hourly and Quarter-Hourly contracts with a spillover multiplier, and show how liquidity conditions affect their strength, direction and timescales.
\end{itemize}

The remainder of this paper is organised as follows. Section~\ref{sec:market} describes the European power market and the mechanics of the IDC market, and presents empirical evidence on order clustering. Section~\ref{sec:model} sets out the SDHP model, the kernel specifications and the estimation method. Section~\ref{sec:results} reports the kernel comparison, the selected specification and the empirical results. Section~\ref{sec:conclusion} concludes and discusses limitations.

\section{European Intraday Continuous Power Markets}
\label{sec:market}
% Short-term trading in Europe is conducted in two main markets: Day-Ahead Market and Intraday market. The day-ahead market is auction based, market participants submit orders in a blind auction where they have no access to the existing orders nor a LOB. For every delivery period, the trading starts 45 days in advance in most EU countries and closes one day before delivery at 12:00 CET. Results are published at
% 12:55 CET. EUPHEMIA is the optimization algorithm responsible for making a single price coupling of European day-ahead electricity markets \cite{nemo2024euphemia}. It simultaneously processes bids/offers, network constraints and cross-border capacities to compute uniform clearing prices per bidding zone. At 14:00 CET the IDC market opens for delivery the next day. 

% Figure \ref{fig:epex_overview} shows the trading timeline at EPEX SPOT. 

% \begin{figure*}[!hbt]
%     \centering
%     \includegraphics[width=.8\textwidth]{Figures/epex_overview (1).png}  
%     \caption{EPEX SPOT power trading timeline overview as for February 2025.}
%     \label{fig:epex_overview}
% \end{figure*}

\begin{figure*}[!hbt]
    \centering
    \includegraphics[width=1\textwidth]{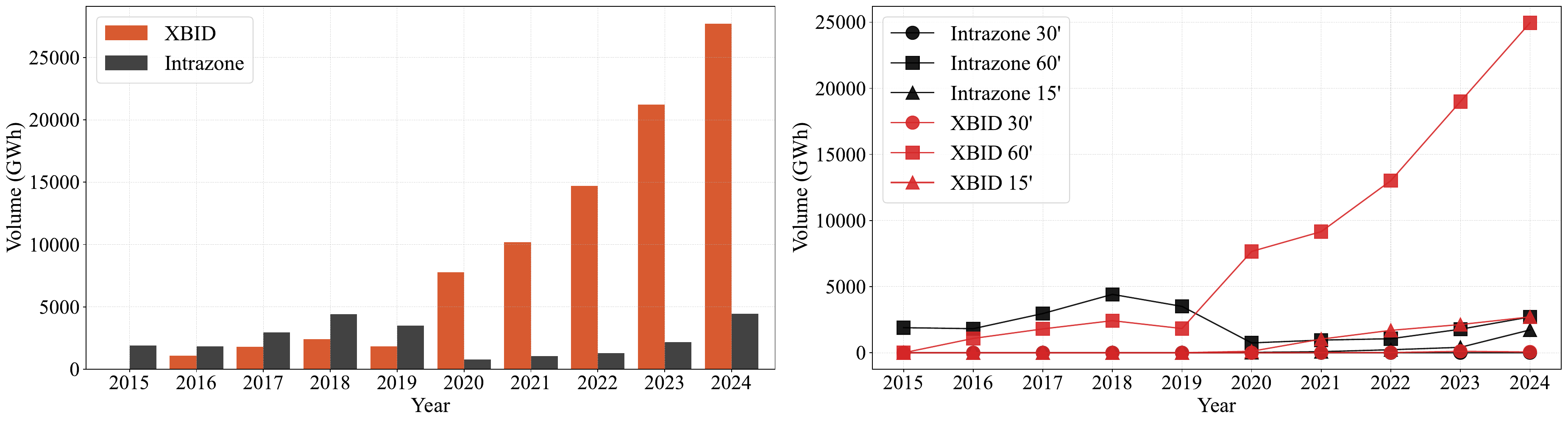}  
    \caption{Annual traded volume across products. Left: Year-by-year evolution of traded volume for XBID and Intrazone product families (GWh) in the Netherlands. Right: Year-by-year evolution of traded volume for each individual product.}
    \label{fig:epex_trends}
\end{figure*}

% \subsection{Intraday Continuous Market}
In the IDC market, market participants can trade several products until shortly before delivery. In 2024, trading for intrazonal contracts remained open until 5 minutes before delivery, while XBID contracts closed 1 hour before delivery. For both intrazonal and XBID markets, several granularities exist: hourly, half-hourly, and quarter-hourly. All these contracts open for trading at 14:00 CET on the day before delivery and can be traded simultaneously. Given the multitude of concurrent contracts available per delivery period, not all of them attract sufficient liquidity to support healthy trading dynamics. Figure \ref{fig:epex_trends} shows the total traded volume per contract type over time in the Netherlands. It reveals the rapid adoption and growing popularity of XBID products, particularly the hourly XBID contract. In fact, in the Netherlands in 2024, the hourly XBID was the most heavily traded contract in terms of the total number of trades, submitted orders, and trading volume, followed by the quarter-hourly XBID products. This suggests that market participants remain dependent on the classical hourly market structure while increasingly utilizing quarter-hourly contracts for finer balancing. This dynamic motivates our choice to study the relationship between these two specific products, which constituted 86\% of the total traded volume in the Netherlands in 2024.

% For every XBID contract, orders are aggregated into a centralized SOB which aggregates all buy (bid) and sell (ask) interests for that specific contract. Every limit order submitted to the book is characterized by a specific direction (buy or sell), a desired volume, and a limit price representing the maximum price a buyer is willing to pay or the minimum a seller is willing to accept. The exchange's matching engine continuously processes these incoming orders based on strict price-time priority. This means that orders offering the most competitive prices—the highest bids and the lowest asks—are prioritized for execution. If multiple orders are submitted at the identical price level, priority is granted to the order that arrived first. 

% To provide participants with granular control over their trading strategies, orders can be submitted with specific execution restrictions. A standard resting limit order carries no restriction, denoted as 'NON'. Alternatively, traders can apply conditions such as 'IOC' (Immediate-or-Cancel), where the order must be executed immediately either fully or partially with any unexecuted portion instantly deleted; 'FOK' (Fill-or-Kill), which mandates immediate execution of the entire volume or complete cancellation; and 'AON' (All-or-None), which requires the order to be executed in its entirety, though it may rest in the book until sufficient opposing liquidity is available.

\subsection{Clustering of Orders}

Point processes are a class of stochastic processes where one realization is defined by a single point in time. The simplest example of a point process is the Poisson point process, in which events occur randomly and independently from each other over time at a constant rate. In a Poisson process, the occurrence of one event has no influence on the likelihood of future events, meaning the process has no memory and does not generate clustering beyond what would arise by chance. In contrast, a Hawkes process is a self-exciting point process in which past events increase the probability of future events occurring shortly afterward. This generates temporal clustering and dependence, as each event can trigger subsequent events, leading to bursts of activity followed by periods of calm.

\begin{figure*}[!hbt]
\centering
\includegraphics[width=\textwidth]{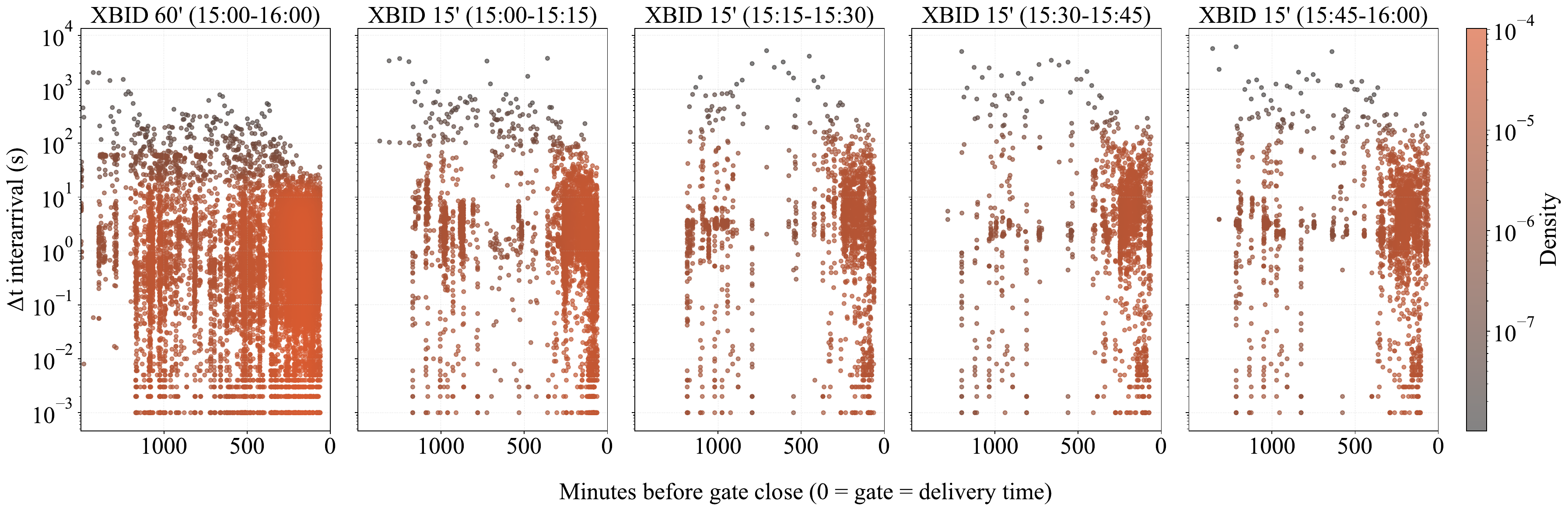}
\caption{Scatter and density plots of order interarrival times against time-to-gate for an XBID hourly contract (2024-01-16 15:00-16:00) and its four constituent XBID quarter-hourly contracts (2024-01-16 15:00-15:15, 15:15-15:30, 15:30-15:45, 15:45-16:00).}
\label{fig:density_kde}
\end{figure*}

Figure ~\ref{fig:density_kde} shows this phenomenon for IDC electricity markets. Rather than being evenly dispersed over time, order arrivals tend to occur in bursts, with periods of intense activity followed by relatively quiet intervals. The color scale represents the density of order arrivals. The figure reveals that high-density regions are concentrated in the minutes immediately preceding delivery, indicating an intensification of trading activity as maturity approaches. This pattern is consistent with the Samuelson effect, which states that volatility and trading intensity tend to increase as a contract approaches its delivery time. This temporal clustering is inconsistent with the independence assumption of the Poisson process but is naturally explained by a self-exciting mechanism in which the arrival of an order increases the likelihood of subsequent orders.

% \subsection{Cancellation time}

% \subsection{Cross product relationship}

\section{Model and Training}
\label{sec:model}

This section provides the full SDHP model. We start by briefly introducing the data and the training setup. We then define the model mathematically. Next, we present the likelihood, its gradients, the stability of the model and the spillover.

\subsection{Data and Training Setup}

We estimate SDHP models following \cite{Morariu-Patrichi04032022} and extending their framework in two key directions. First, we allow the baseline intensities to be state-dependent, enabling the model to capture changes in autonomous order flow across different market regimes. Second, we replace the single exponential decay with a multi-exponential kernel to simultaneously capture the disparate timescales of fast algorithmic reactions and slower, macro-level trading flow.

\begin{figure}[!hbt]
    \centering
    \includegraphics[width=.5\textwidth]{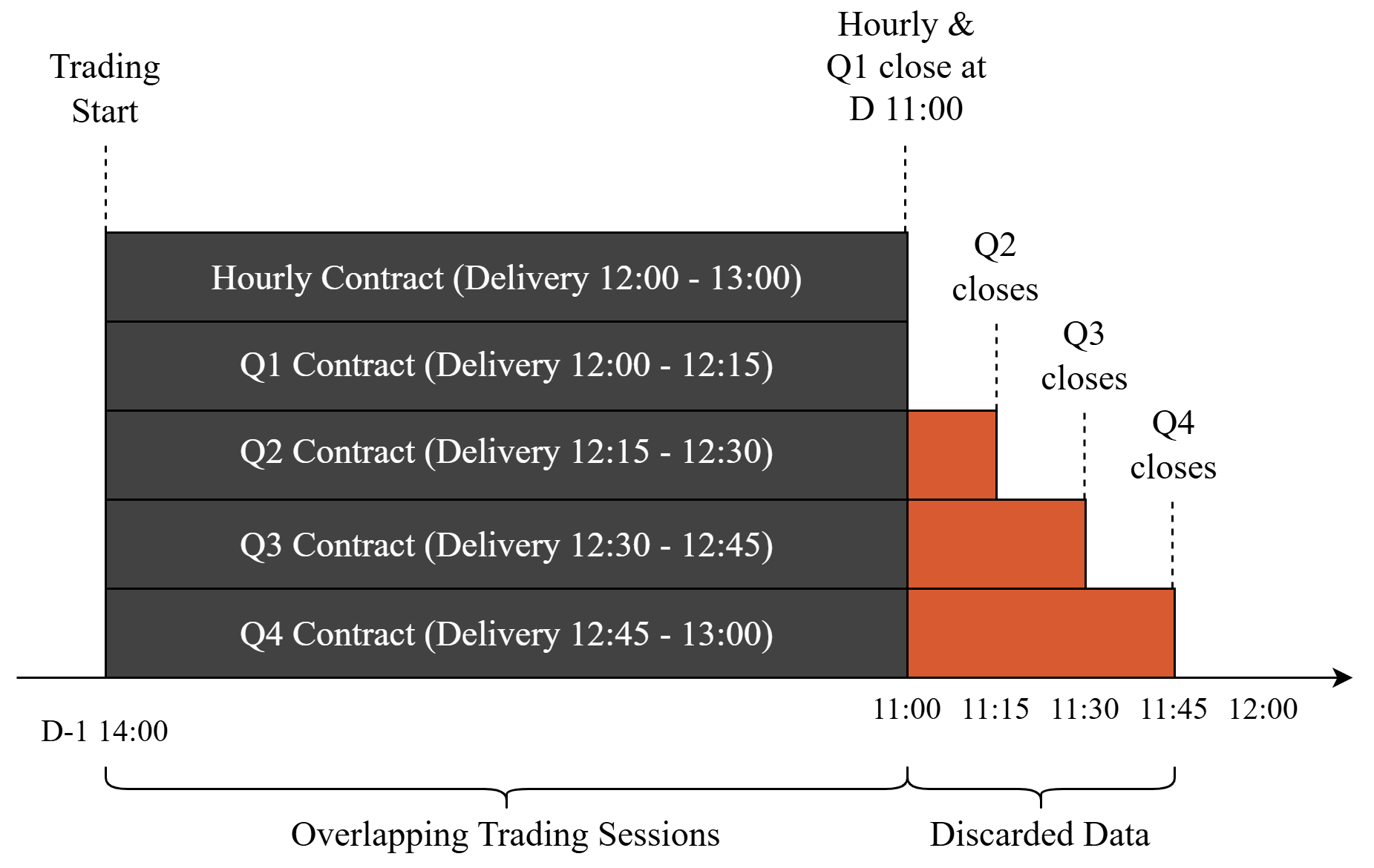}  
    \caption{Temporal alignment of Hourly and Quarter-Hourly contracts. The figure illustrates a single delivery hour (12:00–13:00) alongside its four nested quarter-hour products. The Hourly contract and the first Quarter-Hourly contract (Q1, 12:00–12:15) both close at 11:00. However, Q2 (12:15–12:30) closes at 11:15, Q3 (12:30–12:45) at 11:30, and Q4 (12:45–13:00) at 11:45. Therefore, between 11:00 and 11:45, some contracts are closed while others remain actively traded. Non-overlapping periods (in orange color) are removed to ensure that the Hawkes process is calibrated on purely synchronous LOB dynamics.}
    \label{fig:qh_h}
\end{figure}

For our empirical application, we define a single trading session as the order flow of one specific Hourly XBID contract alongside its four corresponding Quarter-Hourly (Q) constituent contracts spanning the same delivery hour. We focus on the final five hours of trading for every chosen session. However, because XBID contracts close exactly one hour prior to their specific delivery start time, the Gate Closure Times (GCT) of the Hourly and Quarter-Hourly contracts are staggered. To ensure the multivariate Hawkes process is calibrated strictly on synchronous LOB dynamics, we truncate the data to the overlapping active window (illustrated as the grey period in Figure \ref{fig:qh_h}), during which all five contracts are simultaneously open for trading. This truncation is a methodological necessity: including non-overlapping periods where certain contracts are closed would introduce artificial intervals of zero activity. The model would falsely interpret these structural market closures as a genuine lack of liquidity or cross-product interaction, severely biasing the estimation of both the baseline intensities and the cross-excitation kernels. Therefore, we treat each of these precisely defined, fully synchronous sessions as an independent realization.

We fit each delivery contract independently. Our calibration uses contracts from the first month of each quarter of 2024 (January, April, July, and October), which spans seasonal variation in the Dutch IDC market. 

\subsection{The State-dependent Hawkes Process}

To quantify the dynamics between the Hourly (H) and Quarter-Hourly (Q) LOBs under varying market conditions, we model the arrival of limit orders as a multivariate SDHP.

\subsubsection{Event Types and States}
We define a $2$-dimensional event space ($D=2$):
\[
\mathcal{E} = \{H,\, \, Q\}.
\]
Here, $H$ and $Q$ denote the submission of an order in the hourly and quarter-hourly contracts, respectively. 
Simultaneously, we define the state process $S(t)$ using a composite Liquidity Stress Indicator (LSI) constructed from three key microstructural features: the bid–ask spread, mid-price short-term volatility, and queue imbalance.

The mid-price volatility captures short-term price uncertainty and is defined as the rolling standard deviation of the mid-price over the past 10 minutes:
\begin{equation}
\sigma_{\mathrm{mid}}(t)
=
\sqrt{
\frac{1}{N_t}
\sum_{i:\,t_i\in(t-10\text{ min},\,t]}
\left(P_{\mathrm{mid}}(t_i)-\overline{P}_{\mathrm{mid}}(t)\right)^2
},
\label{eq:mid_vol}
\end{equation}
where $\bar{P}_{\mathrm{mid}}(t)$ denotes the average mid-price over the same 10-minute window and $N_t$ is the number of observations in that interval.

Since our objective is to quantify the degree of liquidity imbalance irrespective of its direction, we use the absolute queue imbalance, which measures the magnitude of the liquidity asymmetry between the bid and ask sides, 
\begin{equation}
QI_{\mathrm{abs}}(t)
=
\left|
\frac{\sum_{i \in \mathcal{B}(t)} q_i(t) - \sum_{i \in \mathcal{A}(t)} q_i(t)}
     {\sum_{i \in \mathcal{B}(t)} q_i(t) + \sum_{i \in \mathcal{A}(t)} q_i(t)}
\right|,
\label{eq:qi-abs}
\end{equation}
% where
% \begin{equation}
% V_{\text{bid}}(t) = \sum_{i \in \mathcal{B}(t)} q_i(t), V_{\text{ask}}(t) = \sum_{i \in \mathcal{A}(t)} q_i(t)
% \label{eq:volume}
% \end{equation}

with $\mathcal{B}(t)$ and $\mathcal{A}(t)$ denoting the sets of active bid and ask orders in the LOB at time $t$, respectively, and $q_i(t)$ the volume associated with order $i$. 

The LSI is then defined as
\begin{equation}
\mathrm{LSI}(t)
=
\frac{1}{3}z^{\mathrm{slow}}_{\mathrm{spread}}(t)
+
\frac{1}{3}z^{\mathrm{slow}}_{\mathrm{volatility}}(t)
+
\frac{1}{3}z^{\mathrm{slow}}_{QI_{\mathrm{abs}}}(t).
\label{eq:lsi}
\end{equation}
where $z^{slow}_{\cdot}(t)$ denotes the slow-baseline Z-score of the corresponding feature.

The slow-baseline Z-score is constructed using exponentially weighted estimators for the mean and variance:
\begin{align}
\mu^{slow}_t &= (1-\lambda)\mu^{slow}_{t-1} + \lambda X_t, \\
\sigma^{2,slow}_t &= (1-\lambda)\sigma^{2,slow}_{t-1} + \lambda \big(X_t - \mu^{slow}_t\big)^2, \\
z^{slow}_t &= \frac{X_t - \mu^{slow}_t}{\sigma^{slow}_t},
\label{eq:zslow}
\end{align}
where $X_t$ denotes the feature value and $\lambda$ controls the adaptation speed. We set $\lambda=0.01$, corresponding to an effective memory of approximately 100 observations. 
This slow adaptation helps ensure that persistent liquidity stress remains detectable, rather than being rapidly absorbed into the baseline level, which would otherwise mask sustained regime changes.

The state process is then defined using the empirical distribution of the LSI. Let $q_{0.75}$ denote the $75^{\text{th}}$ percentile of $\mathrm{LSI}(t)$ computed over the full sample. The market state is defined as
\begin{equation}
S(t)
=
\begin{cases}
0, & \mathrm{LSI}(t) < q_{0.75}, \quad \text{(Normal regime)} \\
1, & \mathrm{LSI}(t) \ge q_{0.75}, \quad \text{(Stressed regime)}.
\end{cases}
\label{eq:state-definition}
\end{equation}

The choice of the $75^{\text{th}}$ percentile as the threshold is motivated by two key methodological considerations. First, calibrating a SDHP requires a statistically significant number of events in both regimes to robustly estimate the state-specific baseline intensities and cross-excitation kernels; selecting a higher threshold (e.g., the $95^{\text{th}}$ percentile) would result in data sparsity and noisy parameter estimates for the stressed state. Second, this threshold ensures that the "Stressed" regime represents a genuine, recurring structural state of the market, rather than simply capturing isolated microstructural anomalies or extreme outliers. Finally, this definition ensures that the stressed state corresponds to periods of persistently high volatility and wide bid–ask spreads, while the queue imbalance provides complementary information about the asymmetry of available liquidity between the bid and ask sides of the order book.
% gha mital mn b3d bdl tsswira o indiqi wach avril etc o qualite tl3ha o modify l paragraphe chwiya tas7i7 o kda

% Electricity markets are known to exhibit pronounced intraday seasonality, so one might expect a single global quantile threshold to misrepresent the trading behavior of individual delivery hours. This concern is pre-empted by the construction of the LSI itself. Because each component enters through its slow rolling $z$-score, the raw observables are centered and scaled against their own recent history before aggregation. This removes contract- and time-specific location and dispersion, renders the components dimensionless and comparable, and yields an LSI whose empirical support is stable across delivery periods rather than inheriting the level of intraday activity.

\begin{figure*}[!hbt]
    \centering
    \includegraphics[width=1\textwidth]{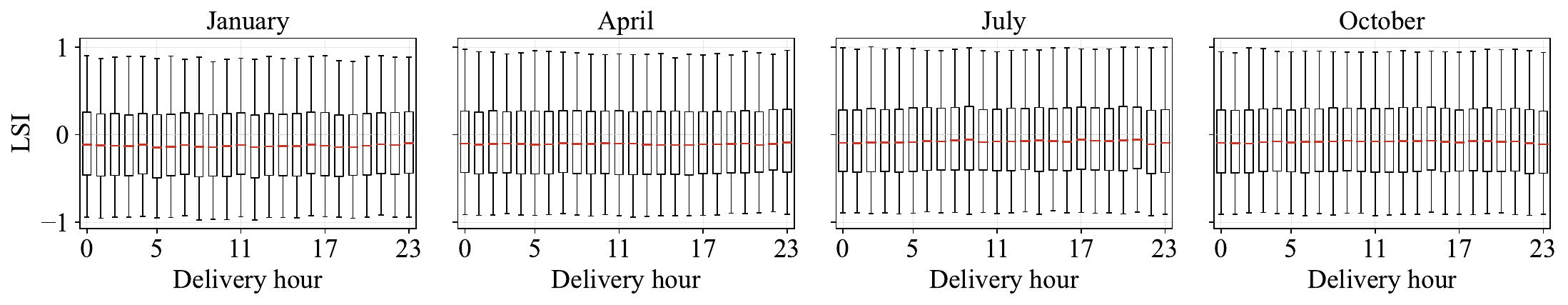}
    \caption{Liquidity Stress Score (LSI) by delivery hour for the first month of each quarter (January, April, July, and October 2024), one panel per month. Each box gives the interquartile range and median of the LSI for contracts of the corresponding delivery hour; whiskers mark the 5th and 95th percentiles. Within each month the boxes are flat and centered across all hours, and the four panels are nearly identical, showing that the slow-$z$ normalization absorbs intraday seasonality and that the LSI distribution is consistent across delivery hours and across the training sample.}
    \label{fig:lss_distr_per_hour}
\end{figure*}

Figure~\ref{fig:lss_distr_per_hour} shows the distribution of the LSI by delivery hour for the first month of each quarter (January, April, July and October 2024). The inter-quartile boxes are centered and essentially flat across all $24$ hours, with no hour displacing its body upward or downward; only the heavy two-sided tails vary mildly. The intraday seasonality has therefore already been absorbed by the normalization and does not survive into the LSI, so a threshold defined on the pooled distribution does not covertly encode time of day.

To confirm that a single fixed threshold transports across sessions, we compare the global $75$th percentile against its per-session counterparts and assess its out-of-sample calibration; results are summarized in
Table~\ref{tab:threshold_calibration}. The per-session thresholds cluster tightly around the global value, and the small median pairwise Wasserstein distance between session distributions indicates that sessions are similar though not identical. We also conducted a leave-one-session-out calibration: for each session we apply the threshold estimated on all remaining sessions and record the realized fraction of stressed events. This fraction is centered on the $25\%$ target and stays within $[0.18,\,0.32]$ across every session. No session collapses to a single regime---even the calmest retains $18\%$ stressed events---so the state variable exercises both regimes in every session and the state-dependent kernels remain identifiable throughout. 

We therefore adopt the global threshold $q^{\text{glob}}_{0.75}=0.264$.

The LSI was introduced in our earlier work \cite{jhabli2026regime}, where we assessed its robustness to the choice of components and threshold levels, as well as the contribution of each component to regime transitions. In that specification, the LSI combined the bid–ask spread, the mid-price volatility and the total depth of the LOB. The robustness checks showed that the sign with which depth entered the index had little effect on the resulting classification, indicating that the component carried limited information about stress. We therefore replace it here with the absolute queue imbalance (Eq.~\ref{eq:qi-abs}), a depth-based quantity with an unambiguous directional interpretation: values close to one indicate a one-sided, fragile book, while values close to zero indicate balanced liquidity on both sides.

\begin{table}[!hbt]
    \centering
    \caption{Transportability of the global stress threshold across sessions.
    Per-session thresholds and pairwise distances are reported relative to the LSI
    standard deviation ($\sigma_{\text{LSI}}$); the leave-one-session-out (LOO)
    stressed fraction applies the threshold estimated on all remaining sessions to
    each held-out session.}
    \vspace{10pt}
    \label{tab:threshold_calibration}
    \small
    \begin{tabular}{lc}
        \toprule
        Quantity & Value \\
        \midrule
        Global threshold $q^{\text{glob}}_{0.75}$        & $0.264$ \\
        Per-session $q_{0.75}$: mean                      & $0.275$ \\
        Per-session $q_{0.75}$: std                       & $0.049$ \;($8.6\%\,\sigma_{\text{LSI}}$) \\
        Per-session $q_{0.75}$: $5$--$95\%$ range         & $[0.191,\,0.350]$ \\
        Median pairwise Wasserstein distance             & $0.060$ \;($10.7\%\,\sigma_{\text{LSI}}$) \\
        \midrule
        LOO stressed fraction: mean                      & $0.256$ \\
        LOO stressed fraction: min / max                 & $0.181\;/\;0.316$ \\
        \bottomrule
    \end{tabular}
\end{table}

\subsection{The State-dependent Hawkes Process}

Let $N(t) = \big(N_{\mathrm{H}}(t),\,N_{\mathrm{Q}}(t)\big)^\top$ be a bivariate counting process, where $N_e(t)$ records the cumulative number of events of type $e\in\mathcal{E}=\{\mathrm{H},\mathrm{Q}\}$ up to time $t$. We model its conditional intensity, $\lambda_e(t)\,dt = \mathbb{E}\!\left[dN_e(t)\mid \mathcal{H}_{t^-}\right]$, as a self-exciting process that adapts to the prevailing market regime. By letting $S(t)\in\{0,1\}$ denote the observed regime (Normal or Stressed, evaluated at $t^-$), we define the state-dependent conditional intensity for target event $e$ directly as:
\begin{equation}
\lambda_e(t)
=
\sum_{x=0}^{1}\mathbf{1}\{S(t)=x\}
\left(
\mu_{x,e}
+
\sum_{e'\in\mathcal{E}}
\sum_{t_i<t:\,e_i=e'}
\kappa^{(x)}_{e'\to e}(t-t_i)
\right).
\label{eq:sdhp_intensity}
\end{equation}
This decomposes the arrival rate into two components:
\begin{itemize}
\item \textbf{Baseline intensity ($\mu_{x,e}\ge 0$):} the exogenous spontaneous rate of arrival for event $e$, depending only on the current regime $x=S(t)$.
\item \textbf{Excitation integral:} the cumulative endogenous impact of all past events of type $e'$ on the current arrival probability of event $e$, modulated by the current regime.
\end{itemize}

The function $\kappa^{(x)}_{e'\to e}(\tau)$ is the \emph{excitation kernel}: it determines how the impact of a past event of type $e'$ on future events of type $e$ decays with the elapsed time $\tau>0$. The specification of $\kappa$ is a central modelling choice of this paper, and we consider four alternatives, introduced next.
% ============================================================
\subsection{Kernel Specifications}
\label{sec:kernels}

Financial microstructure exhibits multi-scale memory: algorithms react in milliseconds, while human traders and slower strategies react over seconds or minutes. A kernel specification must therefore accommodate several coexisting timescales. The following two kernel specification families are considered in this work.

\subsubsection{Exponential-mixture Kernels}

The first family superposes $R$ exponential components,
\begin{equation}
\kappa^{(x)}_{e'\to e}(\tau)
=
\sum_{r=1}^{R}\alpha^{(x,r)}_{e'\to e}\,\exp\!\big(-\beta^{(x,r)}_{e'\to e}\,\tau\big),
\qquad \tau>0,
\label{eq:multi_exp_kernel}
\end{equation}
with magnitudes $\alpha^{(x,r)}_{e'\to e}\ge 0$ and decay rates $\beta^{(x,r)}_{e'\to e}>0$. Each component $r$ isolates one characteristic timescale $1/\beta^{(x,r)}_{e'\to e}$, so that increasing $R$ allows progressively richer structure in the market's memory. We estimate \eqref{eq:multi_exp_kernel} for three alternatives: $R\in\{1,2,3\}$.

\subsubsection{Power-law Kernel}

The second family does not impose a single characteristic decay timescale,
\begin{equation}
\kappa^{(x)}_{e'\to e}(\tau)
=
\alpha^{(x)}_{e'\to e}\,\big(\tau + c^{(x)}_{e'\to e}\big)^{-\eta^{(x)}_{e'\to e}},
\qquad \tau>0,
\label{eq:pow_kernel}
\end{equation}
with $\alpha^{(x)}_{e'\to e}\ge 0$, tail exponent $\eta^{(x)}_{e'\to e}>1$ and regularisation $c^{(x)}_{e'\to e}>0$, the latter bounding the kernel at $\tau\to 0^+$. The constraint $\eta>1$ guarantees an integrable kernel. This specification embodies long-memory dynamics: the excitation decays polynomially rather than geometrically, so remote events retain non-negligible influence.

\subsubsection{Relation between The Two Families}

% Table~\ref{tab:kernels} summarises the structural differences between the two kernel specifications.

% \begin{table}[htbp]
% \centering
% \caption{Kernel specifications compared in this paper.}
% \vspace{10pt}
% \label{tab:kernels}
% \small
% \begin{tabular}{lccc}
% \toprule
% Specification & $\kappa(\tau)$ & $\int_0^\infty\!\kappa(\tau)\,d\tau$ & Memory \\
% \midrule
% Exp-$R$ & $\sum_{r=1}^{R}\alpha_r e^{-\beta_r\tau}$
%         & $\sum_{r=1}^{R}\dfrac{\alpha_r}{\beta_r}$
%         & $R$ running sums \\[6pt]
% Power-law & $\alpha(\tau+c)^{-\eta}$
%         & $\dfrac{\alpha\,c^{\,1-\eta}}{\eta-1}$
%         & full history \\
% \bottomrule
% \end{tabular}
% \end{table}

The two families are more closely related than their functional forms suggest. For $\eta>0$ and $c>0$, the power-law kernel can be written as a weighted average of exponential kernels,
\begin{equation}
(\tau+c)^{-\eta}
=
\frac{1}{\Gamma(\eta)}\int_{0}^{\infty}
\underbrace{u^{\eta-1}e^{-uc}}_{\text{weight } m(u)}
\underbrace{e^{-u\tau}}_{\text{exponential kernel}}\,du,
\label{eq:gamma_mixture}
\end{equation}
in which each decay rate $u$ receives weight $m(u)\propto u^{\eta-1}e^{-uc}$. A power-law kernel is
thus a continuous mixture of exponential kernels with a continuum of decay rates. The $R$-component mixture \eqref{eq:multi_exp_kernel} approximates the same object by
replacing that continuum with $R$ representative decay rates.

Equation~\eqref{eq:gamma_mixture} can be verified directly. It is an instance of Bernstein's theorem \cite{10.1007/BF02592679}, which states that a function on the positive half-line can be written as a mixture of exponentials
precisely when it is completely monotone. The power-law kernel satisfies this condition, so a representation of the form \eqref{eq:gamma_mixture} necessarily exists; the Gamma weight identifies the corresponding mixture. The same representation holds for any completely monotone kernel.

This identity reframes the comparison. Rather than opposing two unrelated functional forms, the choice between them comes back to an empirical question: \emph{how many distinct timescales (i.e., exponential
components) are required to describe the market's memory?}

\begin{figure*}[h]
\centering
\includegraphics[width=\textwidth]{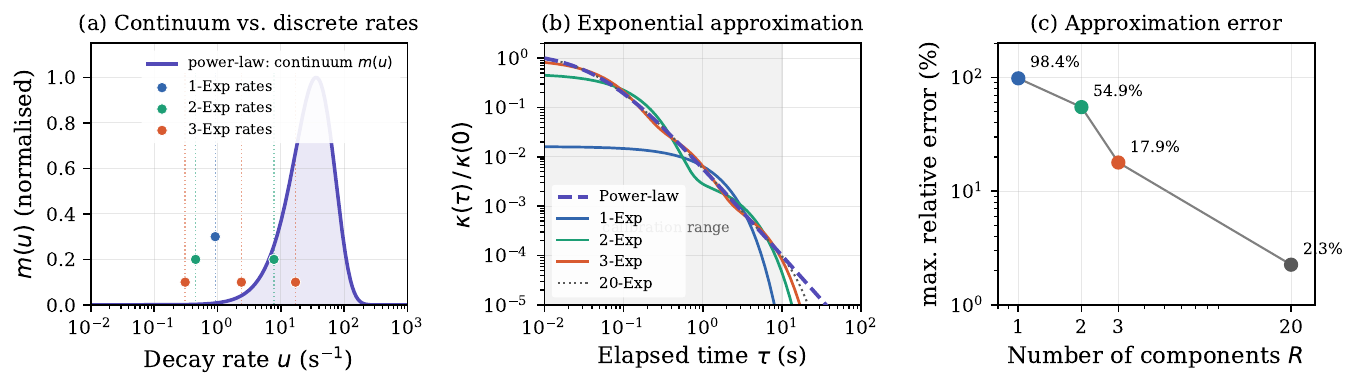}
\caption{Relationship between the exponential-mixture and power-law kernels.
\textbf{(a)} The power-law kernel corresponds to a continuum of exponential decay rates, weighted by the mixing density $m(u)\propto u^{\eta-1}e^{-uc}$ of \eqref{eq:gamma_mixture} (shaded). Markers indicate the decay rates selected for $R$= $1$-, $2$- and $3$ by exponential mixtures fitted to the same kernel; their horizontal positions, not their heights, are informative.
\textbf{(b)} Normalised kernels on a log--log scale. Within the calibration range (shaded) the exponential mixtures track the power-law increasingly closely as $R$ grows; beyond it they decay geometrically while the power-law decays polynomially, so the tail is underestimated for every finite $R$.
\textbf{(c)} Largest relative gap between the $R$-component approximation and the power-law kernel over the calibration range, as a percentage of the latter; it decreases with $R$ but does not vanish. Parameters are illustrative ($\eta=1.8$, $c=0.05$), and the exponential mixtures are fitted to the power-law kernel over $\tau\in[10^{-2},10]$~s.}
\label{fig:kernel_relation}
\end{figure*}

Figure~\ref{fig:kernel_relation} makes this relationship explicit. Panel~(a) contrasts the continuum of decay rates underlying the power-law kernel with the finitely many rates selected by the exponential mixtures. Panel~(b) shows that the approximation is accurate over the range of lags used
for calibration and improves when $R$ increases, while panel~(c) reports the associated approximation error.
Here the error is measured as the largest relative gap between the two kernels over the comparison lag range, expressed as a percentage of the power-law kernel. This represents a worst-case error, and the typical gap is considerably smaller at most lags. The exponential mixtures shown are those that approximate the power-law kernel as closely as possible for a given $R$. Beyond the calibration range, every finite $R$ exponential mixture decays geometrically and therefore underestimates the power-law tail, however large $R$ becomes.

% ============================================================
\subsection{Multi-Session Log-Likelihood}

Let $j=1,\dots,J$ index the independent trading sessions, each with event times and types $\{(t^{(j)}_k, e^{(j)}_k)\}_{k=1}^{K_j}$ observed on the window $[0,T_j]$, alongside an observed state path $S^{(j)}(t)$ evaluated just before each event at $t^{(j)-}_k$. The joint log-likelihood over all sessions is
\begin{equation}
\mathcal{L}(\theta)
=
\sum_{j=1}^{J}
\left[
\sum_{k=1}^{K_j}\log \lambda^{(j)}_{e^{(j)}_k}\!\big(t^{(j)}_k\big)
-
\sum_{e\in\mathcal{E}}\int_0^{T_j} \lambda^{(j)}_{e}(t)\,dt
\right],
\label{eq:LL-consistent}
\end{equation}
where the intensity $\lambda^{(j)}_{e}(t)$ follows \eqref{eq:sdhp_intensity} conditioned on the prevailing regime $S^{(j)}(t^-)$. Expression \eqref{eq:LL-consistent} holds for any kernel; only the
evaluation of $\lambda$ and of the compensator $\int_0^{T_j}\lambda_e$ depends on the specification.

\subsubsection{Computational Structure and History Truncation}
\label{sec:compcost}

The two families differ in the cost of evaluating \eqref{eq:LL-consistent}. For the
exponential kernel, define the partial excitation sums as
\begin{equation}
\Lambda^{(x,r)}_{e'\to e}(t)
=
\sum_{t_i<t:\,e_i=e'}\alpha^{(x,r)}_{e'\to e}\exp\!\big(-\beta^{(x,r)}_{e'\to e}(t-t_i)\big).
\end{equation}
Memorylessness of the exponential gives the exact recursion, for consecutive events separated by $\Delta = t_{k+1}-t_k$,
\begin{equation}
\Lambda^{(x,r)}_{e'\to e}(t_{k+1})
=
e^{-\beta^{(x,r)}_{e'\to e}\Delta}
\Big(\Lambda^{(x,r)}_{e'\to e}(t_{k}) + \alpha^{(x,r)}_{e'\to e}\,\mathbf{1}\{e_k=e'\}\Big).
\label{eq:recursion}
\end{equation}
The intensity therefore depends on the past only through the finitely many running sums $\big(\Lambda^{(x,r)}_{e'\to e}\big)_{r\le R}$: the entire event history is summarised by $R$ numbers per $(e',x,e)$ cell, each updated in constant time as new events arrive. Evaluating
\eqref{eq:LL-consistent} therefore requires a single pass through the data.

The power-law kernel admits no comparable summary. By \eqref{eq:gamma_mixture} it corresponds to a continuum of decay rates, and no finite collection of running sums reproduces it, so $\lambda_e(t)$
depends on the full event history and every event must be compared with all of its predecessors. To keep estimation tractable, we truncate the excitation at a horizon $\Delta_{\max}$, retaining in the sum \eqref{eq:sdhp_intensity} only those past events satisfying $t-t_i\le\Delta_{\max}$. The consequences of this approximation are discussed in Section~\ref{sec:limitations}.

% ============================================================
\subsection{Parameter Count and Computational Cost}

\subsubsection{Parameter Counts}

With $E=2$ event types and $X=2$ regimes, each kernel tensor is indexed by $(e',x,e)$ and therefore carries $E^2X=8$ entries, while the baseline matrix $\mu$ carries $XE=4$. Each additional exponential component thus adds $8$ parameters to $\alpha$ and $8$ to $\beta$. The power-law kernel replaces the decay tensor $\beta$ by the pair $(\eta,c)$, contributing $8$ tail exponents and $8$ regularisation constants. Table~\ref{tab:param_counts} summarizes the number of parameters in $\theta$ for each specification.

\begin{table}[!htbp]
\centering
\caption{Number of estimated parameters per specification ($E=2$, $X=2$).}
\vspace{10pt}
\label{tab:param_counts}
\small
\begin{tabular}{lccccc}
\toprule
\textbf{Model} & $\mu$ & $\alpha$ & $\beta$ & $(\eta,c)$ & \textbf{Total} \\
\midrule
1-Exp     & 4 & 8  & 8  & --      & 20 \\
2-Exp     & 4 & 16 & 16 & --      & 36 \\
3-Exp     & 4 & 24 & 24 & --      & 52 \\
Power-law & 4 & 8  & -- & $8+8$   & 28 \\
\bottomrule
\end{tabular}
\end{table}

The power-law specification is thus \emph{more parsimonious} than the two- and three-component exponential mixtures. It encodes multi-scale memory through a tail exponent rather than through additional timescales. Its cost is computational rather than dimensional.

\subsubsection{Algorithmic Complexity}

The cost of a single likelihood evaluation differs across the two families, as summarised in Table~\ref{tab:complexity}. Exponential kernels admit the recursive update \eqref{eq:recursion}: at each event the running sums $\Lambda_{e',x,e,r}$, a tensor of size $R\,E^2X$, are rescaled by
$e^{-\beta\,\Delta t}$, at a cost of $\mathcal{O}(R E^2 X)$ operations that does not depend on how much history precedes the event. The total cost of one evaluation is therefore $\mathcal{O}(N R E^2 X)$: it grows \emph{linearly} in the number of events $N$, and doubling the sample doubles the cost.

The power-law kernel admits no such recursion, since $(\Delta t + c)^{-\eta}$ cannot be decomposed across consecutive intervals. Each event must instead be compared with every preceding event retained inside the truncation window $\Delta_{\max}$ introduced in Section~\ref{sec:compcost}, at a cost of $\mathcal{O}(\bar{W}E)$ per event, where $\bar{W}=\bar{\lambda}\,\Delta_{\max}$ is the mean number of events falling inside a window and $\bar{\lambda}$ the mean event rate. Two features of this bound matter in practice. First, without truncation $\bar W$ grows with the sample itself and the total cost becomes \emph{quadratic}, $\mathcal{O}(N^2E)$, so that doubling the sample quadruples the cost. Second, $\bar{W}$ is proportional to the local event rate, so the cost per event is not uniform across the sample: it rises with book activity, and is therefore greatest during the dense,
liquid periods that are of primary interest here.

\begin{table}[!htbp]
\centering
\caption{Per-evaluation algorithmic complexity for R-exp and Power law. $N$ denotes the number of events and
$\bar{W}=\bar{\lambda}\Delta_{\max}$ the mean number of events per truncation window.}
\vspace{10pt}
\label{tab:complexity}
\small
\begin{tabular}{lccc}
\toprule
\textbf{Model} & \textbf{Per-event} & \textbf{Total} & \textbf{History} \\
\midrule
$R$-Exp                             & $\mathcal{O}(R E^2 X)$   & $\mathcal{O}(N R E^2 X)$ & Recursive \\
Power-law (no trunc.)           & $\mathcal{O}(N E)$       & $\mathcal{O}(N^2 E)$     & Full \\
Power-law ($\Delta_{\max}$)  & $\mathcal{O}(\bar{W} E)$ & $\mathcal{O}(N \bar{W} E)$ & Windowed \\
\bottomrule
\end{tabular}
\end{table}

% ============================================================
\subsection{Branching Matrix, Stability, and Spillover}
\label{sec:branching}

For each regime $x$, the kernel integrals assemble into the \emph{branching matrix}
$G^{(x)}\in\mathbb{R}_+^{E\times E}$,
\begin{equation}
G^{(x)}_{e'e}
=
\int_{0}^{\infty}\kappa^{(x)}_{e'\to e}(\tau)\,d\tau ,
\label{eq:branching_general}
\end{equation}
whose entry $G^{(x)}_{e'e}$ is the expected number of \emph{direct} offspring events of type $e$ triggered by a single event of type $e'$ while the system is in regime $x$ (rows index the source $e'$, columns the target $e$). Equivalently, it is the total increment in the intensity of $e$ accumulated over all future horizons following one event of type $e'$. Specialising
\eqref{eq:branching_general} to the two families gives
\begin{equation}
G^{(x)}_{e'e}
=
\sum_{r=1}^{R}\frac{\alpha^{(x,r)}_{e'\to e}}{\beta^{(x,r)}_{e'\to e}}
\qquad\text{and}\qquad
G^{(x)}_{e'e}
=
\frac{\alpha^{(x)}_{e'\to e}\,\big(c^{(x)}_{e'\to e}\big)^{1-\eta^{(x)}_{e'\to e}}}
     {\eta^{(x)}_{e'\to e}-1},
\label{eq:branching_both}
\end{equation}
respectively, the latter requiring $\eta^{(x)}_{e'\to e}>1$.

Let $\rho_x=\rho\big(G^{(x)}\big)$ denote the spectral radius of $G^{(x)}$, that is, the largest modulus among its eigenvalues. The stability condition for a multivariate Hawkes process is
\begin{equation}
\rho_x<1,\qquad x\in\{0,1\},
\label{eq:stability}
\end{equation}
which guarantees a \emph{subcritical} process: each event triggers, on average, fewer than one direct successor, so activity initiated by an exogenous shock eventually dies out and the expected number of events it generates is finite. When $\rho_x\ge 1$ the process is supercritical: activity becomes self-sustaining and the expected number of events generated by a single shock is unbounded.

The magnitude of $\rho_x$ has a direct economic interpretation as the \emph{endogeneity}, or reflexivity, of the regime: it measures the share of order flow generated by the market's own past activity rather than by exogenous information arrival. A value near zero describes a market driven
by outside news; a value near one describes a market in which order flow is overwhelmingly a reaction to other order flow, so that shocks propagate with little attenuation.

Under \eqref{eq:stability} the process admits a cluster representation: each exogenous event initiates a cluster of subsequent events, in which every event may in turn trigger further ones. The matrix power $\big(G^{(x)}\big)^{n}$ then records the expected number of events arising at the $n$-th order from the initiating event, and summing over all such rounds yields the
\emph{spillover multiplier}
\begin{equation}
\Psi^{(x)} = \sum_{n=0}^{\infty}\big(G^{(x)}\big)^{n} = \big(I - G^{(x)}\big)^{-1},
\label{eq:spillover}
\end{equation}
whose entry $\Psi^{(x)}_{e'e}$ is the expected total number of events of type $e$ belonging to the cluster initiated by one exogenous event of type $e'$. The $n=0$ term contributes the identity, so the diagonal entries count the initiating event itself; the geometric series converges precisely
when $\rho_x<1$.

% ============================================================
\subsection{Estimation}
\label{subsec:estimation}
We estimate $\theta$ by penalised maximum likelihood. For the exponential family $\theta=\{\mu,\alpha,\beta\}$; for the power-law family $\theta=\{\mu,\alpha,\eta,c\}$.

\subsubsection{Per-state Spectral-radius Penalty}

Because stability strictly requires $\rho_x<1$, we penalise subcriticality during optimisation above a target threshold $\rho^\star<1$:
\begin{equation}
\mathcal{P}(\theta)
=
\sum_{x=0}^{1}\omega_x\,
\Big(\max\{0,\ \rho_x-\rho^\star\}\Big)^2,
\qquad \omega_x>0.
\label{eq:penalty-2states}
\end{equation}
Crucially, $\rho_x$ in \eqref{eq:penalty-2states} is the spectral radius of the \emph{regime-specific} matrix $G^{(x)}$ defined in \eqref{eq:branching_general}, evaluated separately for each regime $x$. The parameters are then estimated by maximising the penalised objective
\begin{equation}
\mathcal{J}(\theta)=\mathcal{L}(\theta)-\mathcal{P}(\theta).
\label{eq:objective}
\end{equation}
We set $\rho^\star=0.95$ and $\omega_x=\omega=10^{3}$ for both regimes. The penalty is a soft constraint: it is inactive wherever $\rho_x\le\rho^\star$, and it grows quadratically beyond the threshold to discourage the supercritical region. Because the constraint binds for some fits, we report the magnitude of $\mathcal{P}(\hat\theta)$ alongside the estimates.

\subsubsection{Parametrisation and Optimisation}
\label{subsubsec:optim}
The objective \eqref{eq:objective} is maximised by the L-BFGS-B quasi-Newton algorithm, using analytical gradients of both $\mathcal{L}$ and $\mathcal{P}$; these are derived in Appendix ~\ref{app:gradients}.

Both families are estimated in logarithmic coordinates. For the exponential family we optimise in $\big(\log\mu,\ \log\alpha,\ \log\beta\big)$, which enforces positivity without explicit constraints
and rescales each gradient by the magnitude of its own parameter, so that baselines and decay rates, which differ by several orders of magnitude, become comparably responsive. For the power-law family we parametrise as $\big(\log\mu,\ \log\alpha,\ \log(\eta-1),\ \log c\big)$; the logarithm of
$\eta-1$ is bounded away from $-\infty$ because the branching integral \eqref{eq:branching_both} is singular as $\eta\downarrow 1$, a singularity that would otherwise render the objective ill-conditioned near that boundary. The initialisation of the decay rates and excitation coefficients, and the box constraints imposed during optimisation, are described in Appendix ~\ref{app:estimation}.

A fit is treated as converged only when the optimiser terminates on its first-order and relative-reduction criteria, rather than on an iteration or function-evaluation budget. Fits terminating on a budget are recorded as non-converged.

% ============================================================
\subsection{Limitations of the Power-Law Approximation}
\label{sec:limitations}

The truncation introduced in Section~\ref{sec:compcost} warrants explicit discussion.

The exponential kernels are evaluated \emph{exactly}: the recursion \eqref{eq:recursion} is algebraically equivalent to the untruncated intensity. The power-law likelihood, by contrast, is evaluated on a kernel truncated at $\Delta_{\max}$, and therefore approximates  different quantity. Two consequences follow.

First, the log-likelihood attained by the power-law specification is an approximation whose bias
decreases with $\Delta_{\max}$ but does not vanish. Any comparison of likelihood-based quantities
across the two families must be read in that light, since one is computed exactly and the other is
not.

Second, the branching integral \eqref{eq:branching_both} for the power-law
integrates the kernel over $[0,\infty)$, whereas the likelihood constrains the kernel only on
$[0,\Delta_{\max}]$. The tail mass beyond the truncation horizon is therefore extrapolated rather
than estimated, and $\rho_x$ inherits that extrapolation---an effect that grows as
$\eta^{(x)}_{e'\to e}\downarrow 1$, precisely where the tail is heaviest. To quantify it, we report
alongside $\rho_x$ the truncated branching radius obtained by integrating $\kappa$ only to
$\Delta_{\max}$; a material gap between the two indicates that the stability estimate is driven by
unobserved tail mass.

These limitations follow from the absence of a recursive representation for the power-law kernel
that makes it computationally heavy, and they are the price of relaxing the assumption of a finite
number of timescales.

\section{Results}
\label{sec:results}

\subsection{Kernel Comparison and Model Selection}
\label{sec:kernel_selection}

We estimate the model separately for each contract, using the first month of
each quarter of 2024: January, April, July and October of the Dutch XBID IDC market from EPEX SPOT. Each contract is fitted with four excitation kernels: the one-, two- and three-component exponential mixtures (1-Exp, 2-Exp, 3-Exp) and the power-law kernel. This yields 2,915 contract-level fits.

This subsection compares the four kernels and selects the specification used in the remainder of the paper. The comparison uses five criteria, reported in the following order. We begin with two formal tests on the time-rescaled residuals: the Kolmogorov--Smirnov test of the marginal distribution and the Ljung--Box test of residual independence. We then report the Wasserstein distance between the empirical residual distribution and its theoretical counterpart, which measures how far a model is from correct specification. We then turn to two likelihood-based criteria, the negative log-likelihood per event and the pairwise probability that one kernel attains a higher likelihood than another on the same contract. Finally, we report the share of estimations that converged, since a specification that fits well but converges rarely is not usable. All criteria are computed per contract and then aggregated, first by month and then across the four months.

\subsubsection{Time-rescaled Residuals}

Goodness of fit is assessed using residual analysis based on the Random Time
Change Theorem. For each event type $e$ we compute the time-rescaled residuals
\begin{equation}
r_{n,e}
=\int_{t_{n-1,e}}^{t_{n,e}} \lambda_e(t)\,dt,
\label{eq:res_plain}
\end{equation}
which are i.i.d.\ $\mathrm{Exp}(1)$ under correct specification. Because the
model is regime-dependent, we also compute state-conditional residuals
\begin{equation}
r^{(x)}_{k,e}
=\int_{t^{(x)}_{k-1,e}}^{t^{(x)}_{k,e}}
\mathbf{1}\{S(t)=x\}\,\lambda_e(t)\,dt,
\label{eq:res_state}
\end{equation}
restricting attention to type-$e$ events occurring in state $x$; these are also $\mathrm{Exp}(1)$ under correct regime-wise specification. Further detail is given in Appendix \ref{app:time_rescaling}.

Equations \eqref{eq:res_plain} and \eqref{eq:res_state} give four residual
streams per contract: Hourly and Quarter-Hourly submissions, each in the Normal and Stressed regimes. Every criterion below is computed on these streams. Two properties of this construction matter for the comparison. First, the null distribution does not depend on the model parameters, so residual-based criteria are directly comparable across the exponential and power-law families, unlike likelihood-based criteria. Second, the residuals test two distinct implications of correct specification, the marginal distribution and independence, and a model may satisfy one while violating the other; we therefore report both.

\subsubsection{Distributional and Independence Tests}

Table~\ref{tab:tests} reports the share of contract-level tests that do not reject the respective null hypotheses: the Kolmogorov--Smirnov test for the $\mathrm{Exp}(1)$ null (left panel) and the Ljung--Box test of no residual serial correlation (right panel).

\begin{table*}[!htbp]
    \centering
    \caption{Share of contract-level tests that do not reject the respective null hypotheses, by month and kernel. The left panel shows the Kolmogorov--Smirnov test for the $\mathrm{Exp}(1)$ null; the right panel shows the Ljung--Box test for the absence of residual serial correlation. Higher is better; the best kernel in each row is in bold.}
    \vspace{10pt}
    \label{tab:tests}
    \small
    \begin{tabular}{@{}lcccc@{\hspace{15pt}}cccc@{}}
    \toprule
    & \multicolumn{4}{c}{\textbf{Kolmogorov--Smirnov (Distribution)}} & \multicolumn{4}{c}{\textbf{Ljung--Box (Independence)}} \\[3pt]
    \cline{2-9} \\[3pt]
    \textbf{Month} & \textbf{1-Exp} & \textbf{2-Exp} & \textbf{3-Exp} & \textbf{Power-law} & \textbf{1-Exp} & \textbf{2-Exp} & \textbf{3-Exp} & \textbf{Power-law} \\ 
    \midrule
    January & 0.826 & 0.910 & \textbf{0.924} & 0.897 & 0.235 & 0.338 & \textbf{0.350} & 0.213 \\
    April   & 0.857 & 0.928 & \textbf{0.931} & 0.886 & 0.245 & 0.255 & \textbf{0.263} & 0.159 \\
    July    & 0.872 & \textbf{0.922} & 0.918 & 0.904 & 0.215 & \textbf{0.249} & 0.237 & 0.117 \\
    October & 0.869 & 0.930 & \textbf{0.932} & 0.914 & 0.190 & \textbf{0.226} & 0.220 & 0.087 \\ 
    \midrule
    All     & 0.856 & 0.922 & \textbf{0.926} & 0.900 & 0.221 & 0.267 & \textbf{0.268} & 0.144 \\
    \bottomrule
    \end{tabular}
\end{table*}

For the Kolmogorov--Smirnov test, the 3-Exp kernel passes most often in three of the four months and pooled across all months, with the 2-Exp kernel close behind and marginally ahead in July. The single-exponential kernel passes least often in every month, and the power-law kernel lies between the single-exponential kernel and the richer mixtures. The ordering is the same in three months out of four.

Pass rates are much lower for the Ljung--Box test than for the Kolmogorov--Smirnov test, between 9\% and 35\% depending on the kernel and month. Residual serial dependence therefore remains for a majority of contracts under all four specifications, and no kernel considered here can be regarded as correctly specified in this respect. The relative ordering is nevertheless consistent: the two- and three-component mixtures leave the least serial dependence and are close to each other, while the power-law kernel leaves clearly the most, passing on only 14.4\% of contract-stream pairs pooled across months.

\subsubsection{Distance to the Theoretical Distribution}

A test reports only whether a contract is rejected, not how large the departure from the null is. Two specifications can therefore show similar pass rates while differing in how closely they fit. We therefore also measure the size of the departure, using the first Wasserstein distance $W_1$ between the empirical distribution of the residuals of a contract and the $\mathrm{Exp}(1)$ law. Smaller values indicate a closer fit. Table~\ref{tab:gof_month} reports the median across contracts by month.

\begin{table}[htbp]
\centering
\caption{Median $W_1$ distance between the time-rescaled residuals and the
$\mathrm{Exp}(1)$ law, by month and kernel. Values are medians across
contracts, averaged over the four event streams and weighted by the number of contracts. Lower is better; the best kernel in each row is in bold. The last column gives the number of contracts.}
\vspace{10pt}
\label{tab:gof_month}
\small
\begin{tabular}{@{}lccccc@{}}
\toprule
\textbf{Month} & \textbf{1-Exp} & \textbf{2-Exp} & \textbf{3-Exp} &
\textbf{Power-law} & \textbf{Contracts} \\ \midrule
January & 0.0857 & 0.0687 & \textbf{0.0657} & 0.0956 & 741 \\
April   & 0.0853 & 0.0680 & \textbf{0.0676} & 0.1013 & 707 \\
July    & 0.0807 & \textbf{0.0694} & 0.0699 & 0.0993 & 738 \\
October & 0.0796 & 0.0594 & \textbf{0.0582} & 0.0892 & 729 \\ \midrule
All     & 0.0834 & 0.0661 & \textbf{0.0651} & 0.0963 & 2,915 \\
\bottomrule
\end{tabular}
\end{table}

The ranking matches that of the two tests and is stable across months. The gain from adding a second exponential component is substantial, from 0.0834 to 0.0661 pooled, a reduction of about 21\%, while the further gain from a third component is small, from 0.0661 to 0.0651, about 1.5\%. The power-law kernel has the largest distance in every month. Absolute fit is somewhat better in October than in the other months for all four kernels, indicating
month-to-month variation in how well the model describes the data, but this
does not alter the ranking.

\begin{figure*}[!htbp]
    \centering
    \includegraphics[width=.9\textwidth]{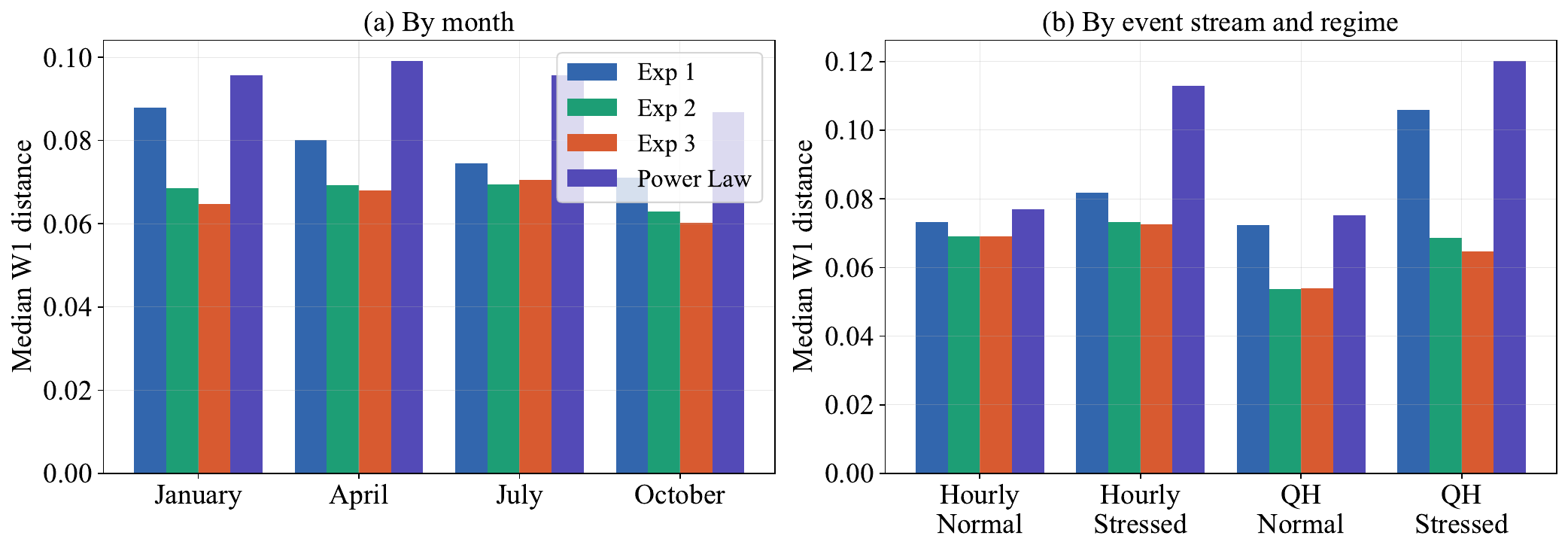}
    \caption{Goodness of fit of the four kernels. Panel (a) reports the median $W_1$ distance between the time-rescaled residuals and the
    $\mathrm{Exp}(1)$ law, by month. Panel (b) reports the same quantity pooled across months and broken down by event stream (Hourly and Quarter-Hourly submissions) and regime (Normal, Stressed). Lower bars indicate a closer fit. The ranking of the kernels is the same in every month, and the differences between kernels are largest in the Stressed regime.}
    \label{fig:kernel_gof}
\end{figure*}

Figure~\ref{fig:kernel_gof} presents the same comparison graphically, together with the breakdown by event stream and regime.

The breakdown by stream in Table~\ref{tab:gof_stream} shows where the kernels differ most.

\begin{table}[htbp]
\centering
\caption{Median $W_1$ distance by event stream and regime, pooled across the
four months. Lower is better; the best kernel in each row is in bold.}
\vspace{10pt}
\label{tab:gof_stream}
\small
\begin{tabular}{@{}llcccc@{}}
\toprule
\textbf{Event} & \textbf{Regime} & \textbf{1-Exp} & \textbf{2-Exp} &
\textbf{3-Exp} & \textbf{Power-law} \\ \midrule
H         & Normal   & 0.0729 & 0.0692 & \textbf{0.0691} & 0.0769 \\
& Stressed & 0.0815 & 0.0735 & \textbf{0.0728} & 0.1128 \\ \midrule
Q & Normal   & 0.0723 & \textbf{0.0538} & 0.0543 & 0.0754 \\
& Stressed & 0.1057 & 0.0685 & \textbf{0.0646} & 0.1202 \\
\bottomrule
\end{tabular}
\end{table}

Differences between kernels are small in the Normal regime and larger in the Stressed regime, where the single-exponential and power-law kernels
deteriorate markedly. The effect is strongest for the Quarter-Hourly stream
under stress, where the distance rises from 0.0646 for the 3-Exp kernel to
0.1057 for the 1-Exp kernel and 0.1202 for the power-law kernel. The kernels
are hardest to distinguish under the least demanding conditions and easiest to distinguish under stress.

\subsubsection{Likelihood-based Comparison}

Table~\ref{tab:nll} reports the negative log-likelihood per event, which
normalises the likelihood by the number of events and therefore allows
comparison across contracts of different lengths, together with the mean
difference in total log-likelihood relative to the 3-Exp kernel on the same
contract.

\begin{table}[htbp]
\centering
\caption{Negative log-likelihood per event and paired log-likelihood
differences relative to the 3-Exp kernel. The paired difference is computed
contract by contract, so both kernels are evaluated on identical data; negative values indicate a worse fit than the reference. Standard deviations across contracts are in parentheses. Lower NLL per event is better.}
\vspace{10pt}
\label{tab:nll}
\small
\begin{tabular}{@{}lccc@{}}
\toprule
\textbf{Kernel} & \textbf{NLL/event} &
\textbf{$\Delta$ LLH vs.\ 3-Exp} & \textbf{Contracts} \\ \midrule
1-Exp     & 1.2708 & $-1072.05$ \; (1097.10) & 2,915 \\
2-Exp     & 1.2067 & $-161.99$ \; (447.99)   & 2,915 \\
3-Exp     & \textbf{1.1950} & --- (reference) & 2,915 \\
Power-law & 1.2480 & $-755.23$ \; (706.52)   & 2,913 \\
\bottomrule
\end{tabular}
\end{table}

The likelihood ranking coincides with the residual-based ranking. The 3-Exp kernel attains the lowest negative log-likelihood per event, followed by the
2-Exp kernel, the power-law kernel and the single-exponential kernel. The paired differences show that the improvement of the 3-Exp over the 2-Exp kernel is small relative to its dispersion across contracts, whereas the gap to the single-exponential and power-law kernels is large. The power-law likelihood is evaluated on a truncated kernel, as discussed in Section~\ref{sec:limitations}, so its likelihood-based figures are not strictly comparable with those of the exponential kernels; the residual-based results above are unaffected by this and give the same ordering.

Table~\ref{tab:winrate} reports the pairwise probability that the kernel in the row attains a higher log-likelihood than the kernel in the column on the same contract. This summary depends only on the sign of the likelihood differences, not their magnitude, and is therefore robust to the small number of contracts with extreme values.

\begin{table}[htbp]
\centering
\caption{Pairwise win rates: $P(\text{row kernel beats column kernel})$, computed contract by contract and pooled across the four months. A value above 0.5 indicates that the row kernel fits better on a majority of contracts.}
\vspace{10pt}
\label{tab:winrate}
\small
\begin{tabular}{@{}lcccc@{}}
\toprule
 & \textbf{1-Exp} & \textbf{2-Exp} & \textbf{3-Exp} & \textbf{Power-law} \\ \midrule
\textbf{1-Exp}     & ---   & 0.010 & 0.003 & 0.233 \\
\textbf{2-Exp}     & 0.990 & ---   & 0.175 & 0.982 \\
\textbf{3-Exp}     & 0.997 & 0.825 & ---   & 0.995 \\
\textbf{Power-law} & 0.767 & 0.018 & 0.005 & ---   \\
\bottomrule
\end{tabular}
\end{table}

The win rates confirm a clear overall ordering. The 3-Exp kernel fits better than the single-exponential kernel on 99.7\% of contracts, better than the power-law kernel on 99.5\%, and better than the 2-Exp kernel on 82.5\%. The power-law kernel beats the single-exponential kernel on 76.7\% of contracts but is beaten by both richer exponential mixtures of more than 98\%. The only comparison that is not close to unanimous is between the two- and three-component mixtures, consistent with the small difference in their residual statistics.

\subsubsection{Convergence}

Table~\ref{tab:convergence} reports the number and share of contract-level estimations satisfying the convergence criterion stated in Section~\ref{subsubsec:optim}.

\begin{table}[htbp]
\centering
\caption{Convergence of the contract-level estimations by kernel.}
\vspace{10pt}
\label{tab:convergence}
\small
\begin{tabular}{@{}lccc@{}}
\toprule
\textbf{Kernel} & \textbf{Attempted} & \textbf{Converged} &
\textbf{Share converged} \\ \midrule
1-Exp     & 2,915 & 2,901 & 99.5\% \\
2-Exp     & 2,915 & 2,860 & 98.1\% \\
3-Exp     & 2,915 & 2,788 & 95.6\% \\
Power-law & 2,915 & 2,402 & 82.4\% \\
\bottomrule
\end{tabular}
\end{table}

Convergence rates decline as the kernel becomes richer, which is expected since both the number of parameters and the difficulty of the optimisation increase. The power-law kernel converges least often, reflecting the poorer conditioning of its parameterisation near the boundary $\eta \to 1$. Convergence should be read alongside the fit statistics, because reported estimates from a fit that stopped on a budget depend on where the optimiser halted rather than on the data.

\subsubsection{Model Selection}

The five criteria agree. The three-component exponential mixture attains the highest pass rate on both residual tests, the lowest median $W_1$ distance in three of the four months and pooled across all months, the lowest negative log-likelihood per event, and a win rate above 0.82 against every competing kernel. The single-exponential kernel is dominated on every criterion, which indicates that a single timescale is too restrictive for this market. The power-law kernel is also dominated: it fits worse than both richer exponential mixtures on the residual criteria, which are the criteria comparable across the two families, it leaves the most residual serial dependence, and it converges least often.

The comparison between the two- and three-component mixtures is closer. The 3-Exp kernel is better on all pooled criteria, but the margins are small: the median $W_1$ distance differs by about 1.5\%, the residual test pass rates by less than half a percentage point, and the 2-Exp kernel is marginally better in July on all three residual criteria. We nevertheless select the 3-Exp kernel for two reasons. First, it is better or equal on every pooled criterion and in three of the four months, so the choice does not rest on a single period. Second, the difference between the two specifications is largest in the Stressed regime and for the Quarter-Hourly stream, which are the conditions of primary interest in this paper; restricting the kernel to two components would sacrifice fits precisely where the model is used most.

The remainder of the results are based on the three-component exponential kernel. 

\subsection{Exogenous and Endogenous Order Flow}
\label{sec:endogeneity}

We now interpret the estimated three-exponential specification. Of the 2,915 contract-level estimations, 2,788 (95.6\%) satisfy the convergence criterion and are subcritical in both regimes; these form the sample analysed below. All statistics are computed per contract and then aggregated. Every comparison between regimes is made \emph{within} contract, so each contract is compared only with itself. Contracts differ widely in how active they are, depending on the delivery hour and the day of the week, and comparing regimes within a contract removes those differences from the comparison.

\subsubsection{Baseline Intensities}

The baseline intensity $\mu_{x,e}$ measures the rate of order arrivals that the model attributes to sources outside the order book: fundamental information, scheduled position adjustments, and forecast revisions. Table~\ref{tab:mu} reports its estimated value by regime and contract.

\begin{table*}[htbp]
\centering
\caption{Baseline intensities $\mu_{x,e}$ by contract, regime and month, in events per second. Values are medians across contracts, with the interquartile range in parentheses. ``Ratio S/N'' is the median of the within-contract ratio $\mu_{\mathrm{Stressed}}/\mu_{\mathrm{Normal}}$ and ``Share $>1$'' the fraction of contracts for which that ratio exceeds one; both isolate the regime effect from differences in activity across contracts.}
\vspace{10pt}
\label{tab:mu}
\small
\begin{tabular}{@{}llcccc@{}}
\toprule
\textbf{Month} & \textbf{Contract} & \textbf{Normal} & \textbf{Stressed} &
\textbf{Ratio S/N} & \textbf{Share $>1$} \\ \midrule
January & H & 0.0355 (0.0431) & 0.0305 (0.0446) & 0.830 & 0.359 \\
                         & Q & 0.0146 (0.0194) & 0.0178 (0.0268) & 1.203 & 0.583 \\ \midrule
April   & H & 0.0253 (0.0276) & 0.0224 (0.0312) & 0.856 & 0.379 \\
                         & Q & 0.0099 (0.0155) & 0.0122 (0.0219) & 1.090 & 0.538 \\ \midrule
July    & H & 0.0251 (0.0216) & 0.0209 (0.0259) & 0.834 & 0.374 \\
                         & Q & 0.0179 (0.0222) & 0.0234 (0.0323) & 1.201 & 0.607 \\ \midrule
October & H & 0.0353 (0.0309) & 0.0357 (0.0385) & 0.974 & 0.473 \\
                         & Q & 0.0294 (0.0308) & 0.0342 (0.0381) & 1.153 & 0.598 \\ \midrule
All     & H & 0.0300 (0.0309) & 0.0268 (0.0354) & 0.875 & 0.397 \\
                         & Q & 0.0170 (0.0235) & 0.0208 (0.0321) & 1.169 & 0.582 \\
\bottomrule
\end{tabular}
\end{table*}

\begin{figure*}[!htbp]
    \centering
    \includegraphics[width=\textwidth]{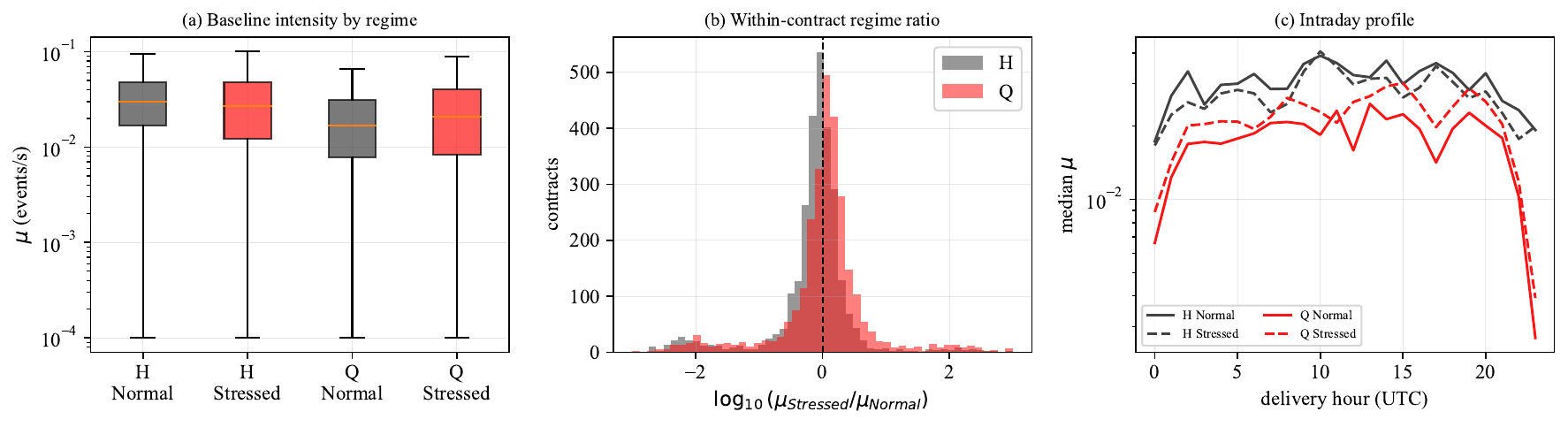}
    \caption{Baseline intensities. Panel (a) shows the distribution across contracts of $\mu$ for each contract and regime, on a logarithmic scale.
    Panel (b) shows the distribution of the within-contract ratio $\mu_{\mathrm{Stressed}}/\mu_{\mathrm{Normal}}$; mass to the right of zero indicates a higher exogenous arrival rate under stress. Panel (c) shows the median baseline by delivery hour, separately for each contract and regime.}
    \label{fig:mu}
\end{figure*}

Figure~\ref{fig:mu} shows the distribution of the baseline across contracts, the distribution of the within-contract regime ratio, and the intraday profile of the baseline by delivery hour.

The intraday profile in panel (c) is the clearest result of this subsection. Exogenous order flow follows a pronounced daily cycle: it is lowest for
deliveries in the early morning hours, rises through the morning to a plateau between the ninth and seventeenth delivery hour, and falls for deliveries late in the evening, with the median baseline for the Quarter-Hourly contract varying by roughly an order of magnitude across the
day. This is the shape one would expect if the baseline is capturing genuine outside information since imbalance in an intraday power market is driven by the same daily cycle of load and vRES generation. The two contracts differ in level throughout: the Hourly baseline lies above the Quarter-Hourly baseline at every delivery hour, consistent with the Hourly contract being the principal venue for exogenous position adjustment.

The regime effect on the baseline is contract-specific, and the two contracts move in opposite directions. The median within-contract ratio is 0.875 for the Hourly contract and 1.169 for the Quarter-Hourly contract, and the share of contracts with a ratio above one is 0.397 and 0.582, respectively. The pattern holds in all four months. Stress therefore does not uniformly raise the rate of exogenous order arrivals: exogenous activity shifts \emph{away} from the principal contract and \emph{towards} the granular one. This is the first indication of a broader pattern documented below, in which the Quarter-Hourly contract gains relative importance under stress.

The relative weight of the two contracts also varies seasonally. In the Normal regime the Hourly baseline exceeds the Quarter-Hourly baseline by a factor of 2.4 in January and 2.6 in April, but only 1.4 in July and 1.2 in October, so the granular contract accounts for a substantially larger share of exogenous order flow in the second half of the year.

Two features of panel (a) qualify these numbers. The lower whisker of every box reaches $10^{-4}$, which is the lower bound imposed on $\mu$ during
optimisation, so for a subset of contracts the baseline is not interior. The same feature is visible in panel (b) as a concentration of mass at a ratio of exactly one, arising from contracts in which the baseline is at the bound in both regimes. The baseline should therefore be read as an upper bound on the exogenous share for those contracts, and the aggregate statement that endogeneity is high is, if anything, conservative.

\subsubsection{The Branching Matrix and Degree of Endogeneity}

Table~\ref{tab:branching} reports the estimated branching matrix. Entry
$G^{(x)}_{e'e}$ is the expected number of type-$e$ orders directly triggered by one type-$e'$ order in regime $x$.

\begin{table}[htbp]
\centering
\caption{Direct branching matrix $G^{(x)}$ and spectral radius $\rho_x$ by regime. Values are medians across the 2,787 contracts. Rows index the source contract, columns the target. The final column reports $\rho_x$, which equals the share of order flow generated endogenously by the market's own past.}
\vspace{10pt}
\label{tab:branching}
\small
\begin{tabular}{@{}llcccc@{}}
\toprule
\textbf{Regime} & \textbf{Source} & \textbf{$\rightarrow$ H} &
\textbf{$\rightarrow$ Q} & & \textbf{$\rho_x$} \\ \midrule
Normal  & H & 0.8715 & 0.0212 & & 0.9518 \\
                          & Q & 0.0524 & 0.9047 & & \\ \midrule
Stressed & H & 0.8777 & 0.0349 & & 0.9555 \\
                          & Q & 0.0490 & 0.8574 & & \\
\bottomrule
\end{tabular}
\end{table}

Two features stand out. First, the matrix is strongly diagonally dominant. An Hourly order triggers on average 0.87 further Hourly orders but only 0.02
Quarter-Hourly orders, and a Quarter-Hourly order triggers 0.90 further Quarter-Hourly orders but only 0.05 Hourly orders. Direct self-excitation is
therefore between seventeen and forty-one times larger than direct cross-excitation. Order flow in each contract is overwhelmingly a reaction to activity in the same contract: order splitting, queue competition and quote revision within a single book.

Second, and more importantly, the spectral radius is close to unity in both regimes, at 0.952 under Normal conditions and 0.956 under Stress. Since $\rho_x$ is the share of activity generated endogenously, roughly 95\% of observed order flow is a reaction to previous order flow, and only about 5\% is attributable to information arriving from outside the book. The intraday power market is therefore highly reflexive: almost all of what happens in the order book is the market responding to itself. This near-critical value also means that the market sits close to the boundary at which shocks cease to be damped, which is the sense in which liquidity in this market is fragile.

The regime effect on endogeneity is small. Table~\ref{tab:paired} reports the median within-contract difference between regimes, with bootstrap confidence intervals and the share of contracts for which the difference is positive.

\begin{table*}[htbp]
\centering
\caption{Within-contract differences between the Stressed and Normal regimes. Each row reports the median of $q_{\mathrm{Stressed}}-q_{\mathrm{Normal}}$ computed contract by contract, a 95\% percentile bootstrap confidence interval for that median, and the share of contracts with a positive difference. A confidence interval excluding zero indicates a regime effect that is consistent across contracts.}
\vspace{10pt}
\label{tab:paired}
\small
\begin{tabular}{@{}llcccc@{}}
\toprule
\textbf{Quantity} & \textbf{Month} & \textbf{Median diff.} &
\textbf{95\% CI} & \textbf{Share $>0$} & \textbf{$n$} \\ \midrule
$\rho$
 & January & $+0.0031$ & $[+0.0009,\,+0.0062]$ & 0.542 & 690 \\
 & April   & $-0.0012$ & $[-0.0042,\,+0.0010]$ & 0.478 & 663 \\
 & July    & $+0.0010$ & $[-0.0022,\,+0.0041]$ & 0.514 & 722 \\
 & October & $+0.0086$ & $[+0.0052,\,+0.0127]$ & 0.603 & 712 \\
 & All     & $+0.0025$ & $[+0.0012,\,+0.0040]$ & 0.535 & 2{,}787 \\ \midrule
$\Psi_{H\to Q}$
 & January & $+0.927$ & $[+0.562,\,+1.390]$ & 0.613 & 690 \\
 & April   & $+0.869$ & $[+0.518,\,+1.273]$ & 0.572 & 663 \\
 & July    & $+0.054$ & $[-0.278,\,+0.437]$ & 0.507 & 722 \\
 & October & $+0.327$ & $[+0.071,\,+0.535]$ & 0.545 & 712 \\
 & All     & $+0.542$ & $[+0.381,\,+0.720]$ & 0.558 & 2{,}787 \\ \midrule
$\Psi_{Q\to H}$
 & January & $-1.502$ & $[-2.328,\,-0.613]$ & 0.426 & 690 \\
 & April   & $-4.227$ & $[-6.052,\,-2.649]$ & 0.371 & 663 \\
 & July    & $-0.619$ & $[-1.291,\,+0.182]$ & 0.468 & 722 \\
 & October & $+0.012$ & $[-0.792,\,+0.603]$ & 0.501 & 712 \\
 & All     & $-1.233$ & $[-1.624,\,-0.756]$ & 0.443 & 2{,}787 \\ \midrule
Asymmetry $A$
 & January & $+0.219$ & $[+0.135,\,+0.320]$ & 0.628 & 690 \\
 & April   & $+0.273$ & $[+0.158,\,+0.429]$ & 0.618 & 663 \\
 & July    & $+0.083$ & $[+0.000,\,+0.143]$ & 0.536 & 722 \\
 & October & $+0.077$ & $[+0.007,\,+0.176]$ & 0.546 & 712 \\
 & All     & $+0.148$ & $[+0.126,\,+0.201]$ & 0.581 & 2{,}787 \\
\bottomrule
\end{tabular}
\end{table*}

Pooled across months, endogeneity is higher under stress, but the difference is $+0.0025$ and the share of contracts with a positive difference is only 0.535, barely above one half. The effect is also not stable across months: it is positive and significant in January and October, indistinguishable from zero in July, and negative in point estimate in April. We therefore do not read the data as showing that market stress raises the degree of endogeneity. What the results do show is that endogeneity is uniformly high: the market is close to critical in both regimes, and the Stressed regime is not distinguished by being more reflexive. As shown next, what stress changes is not how much of the order flow is endogenous, but where it flows.

\subsection{Cross-Product Liquidity Spillovers}
\label{sec:spillovers}

\subsubsection{Direct and Total Transmission}

The direct entries of $G^{(x)}$ understate cross-product transmission, because an order that triggers an order in the other contract also triggers further orders in its own contract, each of which can spill over in turn. The total effect is given by the spillover multiplier $\Psi^{(x)}=(I-G^{(x)})^{-1}$ of \eqref{eq:spillover}. Table~\ref{tab:spillover} reports both.

\begin{table}[htbp]
\centering
\caption{First-round and total cross-product transmission by regime. $G$ is the direct branching entry, which counts the orders triggered in the first round; $\Psi$ is the corresponding entry of the spillover multiplier, which counts the orders appearing in the whole cluster initiated by one exogenous order. ``Amplif.'' is the median across contracts of the per-contract ratio $\Psi/G$, not the ratio of the medians reported in the adjacent columns. The asymmetry index is $A=(\Psi_{H\to Q}-\Psi_{Q\to H})/(\Psi_{H\to Q}+\Psi_{Q\to H})$; a negative value indicates that total transmission runs predominantly from the Quarter-Hourly to the Hourly contract. All values are medians across contracts.}
\vspace{10pt}
\label{tab:spillover}
\small
\begin{tabular}{@{}lcccccc@{}}
\toprule
& \multicolumn{3}{c}{$H \rightarrow Q$} & \multicolumn{2}{c}{$Q \rightarrow H$} & \\
\cmidrule(lr){2-4}\cmidrule(lr){5-6}
\textbf{Regime} & $G$ & $\Psi$ & Amplif. & $G$ & $\Psi$ &
\textbf{$A$} \\ \midrule
Normal   & 0.021 & 2.390 & $\times$113 & 0.052 & 6.675 & $-0.442$ \\
Stressed & 0.035 & 3.393 & $\times$97  & 0.049 & 5.238 & $-0.175$ \\
\bottomrule
\end{tabular}
\end{table}

The two quantities in Table~\ref{tab:spillover} measure different things. The direct entry $G^{(x)}_{e'e}$ counts the type-$e$ orders triggered by one type-$e'$ order in the first round only. The multiplier entry $\Psi^{(x)}_{e'e}$ counts the type-$e$ orders appearing in the entire cluster initiated by one exogenous type-$e'$ order, summing the first round and every subsequent one. Both are expressed in the same units, expected orders per initiating order, so their ratio is well defined: it is the number of orders generated across all rounds divided by the number generated in the first.

That ratio is large. For transmission from the Hourly to the Quarter-Hourly contract, the median across contracts is 112 in the Normal regime and 97 under stress. The reason is the near-critical diagonal documented above. An order that crosses into the other book arrives in a book that multiplies its own order flow by a factor of roughly $1/(1-0.9)$, and each of the resulting orders can cross back. A weak first-round link therefore produces a large cumulative effect. Cross-product spillovers in this market are potentially important even though the first-round coefficients appear negligible, and a measure based on $G^{(x)}$ alone would understate them by about two orders of magnitude.

Because $\Psi^{(x)}$ is a non-linear function of $G^{(x)}$, the median of $\Psi^{(x)}$ across contracts is not the multiplier implied by the median $G^{(x)}$. The two blocks of Table~\ref{tab:spillover} are therefore computed separately and should not be chained; the amplification column reports the median of the per-contract ratios rather than the ratio of the medians.

\subsubsection{The Direction of Transmission and Its Regime Dependence}

In both regimes the asymmetry index is negative: total transmission runs predominantly from the Quarter-Hourly to the Hourly contract. Under Normal conditions one exogenous Quarter-Hourly order ultimately generates 6.68 Hourly orders over the cluster it initiates, whereas one exogenous Hourly order generates 2.39 Quarter-Hourly orders, a ratio of 2.8 to 1. The more granular contract leads the coarser one. This is consistent with the Quarter-Hourly contract being where imbalance is expressed first, with the Hourly contract subsequently repriced or hedged against it.

Market stress compresses this asymmetry. Under Stress the ratio falls from 2.8 to 1.5, and the asymmetry index moves from $-0.442$ to $-0.175$. The within-contract results in Table~\ref{tab:paired} show that this is the most robust regime effect in the estimates: the asymmetry increases under stress in all four months, every confidence interval excludes zero, and the difference is positive for 58.1\% of contracts. The two components move in opposite directions, as they should if the change is a genuine reorientation of the market rather than a level effect: $\Psi_{H\to Q}$ rises by 0.54 and $\Psi_{Q\to H}$ falls by 1.23, both significantly pooled across months. The direct coefficient $G_{H\to Q}$ rises by 64\% under stress, from 0.021 to 0.035, while $G_{Q\to H}$ falls slightly.

The economic reading is that the leadership structure of the market weakens under stress. In calm conditions the two books are linked but hierarchical, with the Quarter-Hourly contract setting the pace. Under stress the Hourly contract becomes a materially stronger source of spillover and the two contracts influence each other almost equally. For a participant hedging across the two contracts, this means the direction in which liquidity shocks propagate is not a stable property of the market: a hedge calibrated on the leadership structure of calm periods becomes less reliable precisely when it is most needed.

\subsubsection{Intraday and Weekly Variation}

\begin{figure*}[!htbp]
    \centering
    \includegraphics[width=\textwidth]{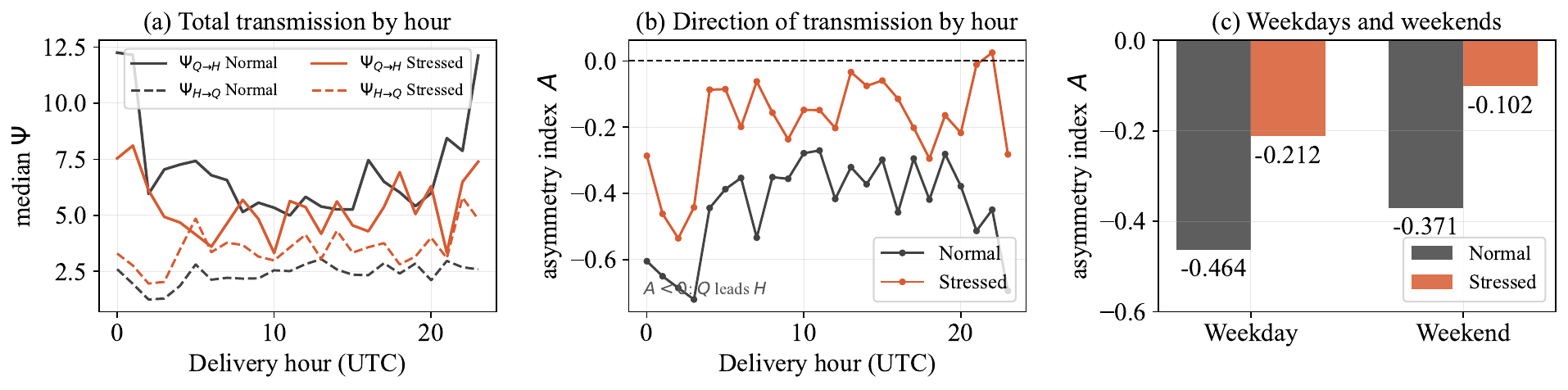}
    \caption{Variation of cross-product spillovers over the delivery cycle and the week. Panel (a) shows the median total transmission $\Psi$ across all eligible contracts from the four estimation months, grouped by delivery hour for both directions and both regimes; the vertical distance between the two lines of a given regime is the imbalance between the directions. Panel (b) shows the median asymmetry index $A=(\Psi_{H\to Q}-\Psi_{Q\to H})/(\Psi_{H\to Q}+\Psi_{Q\to H})$ by delivery hour. The index is bounded between $-1$ and $1$ and is scale-free: $A=0$ means the two contracts transmit equally, $A<0$ means transmission runs predominantly from the Quarter-Hourly to the Hourly contract, and $A>0$ the reverse. Panel (c) compares the asymmetry index on weekdays and at weekends.}
    \label{fig:spillovers}
\end{figure*}

Figure~\ref{fig:spillovers} shows how cross-product transmission varies over the delivery cycle and between weekdays and weekends.

The dominance of the Quarter-Hourly contract is strongly time-varying. Under Normal conditions the asymmetry index is most negative for deliveries in the overnight hours, reaching $-0.72$ at hour 3 and $-0.69$ at hour 23, and is weakest around midday, at $-0.27$ at hour 11. The total transmission
$\Psi_{Q\to H}$ is correspondingly larger overnight, at 12.2 for deliveries at hours 0, 1 and 23, against approximately 5 for midday deliveries. Overnight delivery hours are thin, and in a thin market the Hourly book appears to depend heavily on information expressed first in the Quarter-Hourly contract. During the middle of the day, when both books are more actively traded, the two contracts are closer to equal footing.

Under Stress the intraday pattern flattens: the asymmetry index remains within a narrower band and is close to zero for most delivery hours, and for deliveries at hour 22 it is marginally positive. Panel (b) also shows that the regime effect holds at every point of the delivery cycle: the Stressed curve lies above the Normal curve for all twenty-four delivery hours, so the compression of the asymmetry under stress is not driven by any particular part of the day. The regime effect and the intraday effect therefore work in the same direction, and the strongest hierarchical structure occurs where both point that way, in calm overnight markets.

Differences between weekdays and weekends are modest by comparison. The asymmetry index is $-0.464$ on weekdays and $-0.371$ at weekends in the Normal
regime, and $-0.212$ against $-0.102$ under Stress, so weekends are slightly more symmetric in both regimes but the effect is small relative to the intraday and regime variation.

\subsubsection{Seasonal Variation}

The four estimation months span the seasonal cycle of the Dutch power system: January represents winter demand with limited solar generation, 
April the spring period with rapidly increasing solar generation and variable wind conditions, July the summer solar maximum, and October the autumn transition. Table~\ref{tab:seasonal} reports the main quantities by month and regime.

\begin{table}[htbp]
\centering
\caption{Endogeneity and cross-product spillovers by month and regime. All values are medians across contracts. The final row of each block reports the range across the four months, which measures how much the market's structure varies over the seasonal cycle.}
\vspace{10pt}
\label{tab:seasonal}
\small
\begin{tabular}{@{}lccccc@{}}
\toprule
\textbf{Month} & $\rho$ & $\Psi_{H\to Q}$ & $\Psi_{Q\to H}$ &
\textbf{Asymmetry $A$} & $n$ \\ \midrule
\multicolumn{6}{l}{\emph{Normal regime}} \\
January & 0.9536 & 1.96 & 7.21  & $-0.541$ & 690 \\
April   & 0.9596 & 2.42 & 12.31 & $-0.659$ & 663 \\
July    & 0.9510 & 3.03 & 6.04  & $-0.298$ & 722 \\
October & 0.9368 & 1.95 & 4.64  & $-0.347$ & 712 \\
\emph{Range} & \emph{0.023} & \emph{1.08} & \emph{7.67} & \emph{0.361} & \\ \midrule
\multicolumn{6}{l}{\emph{Stressed regime}} \\
January & 0.9578 & 3.78 & 5.20 & $-0.165$ & 690 \\
April   & 0.9593 & 4.06 & 6.03 & $-0.207$ & 663 \\
July    & 0.9539 & 3.44 & 5.22 & $-0.160$ & 722 \\
October & 0.9521 & 2.60 & 4.85 & $-0.189$ & 712 \\
\emph{Range} & \emph{0.007} & \emph{1.46} & \emph{1.18} & \emph{0.047} & \\
\bottomrule
\end{tabular}
\end{table}

The market's structure varies considerably across the seasonal cycle in calm conditions. The asymmetry index ranges from $-0.659$ in April to $-0.298$ in July, and the total transmission from the Quarter-Hourly to the Hourly contract ranges from 12.31 in April to 4.64 in October, a factor of 2.7. April is the most extreme month on every measure: it has the highest endogeneity, the strongest Quarter-Hourly dominance and by far the largest $\Psi_{Q\to H}$. This seasonal pattern is consistent with the variability of vRES generation in spring, when solar generation is increasing rapidly while wind conditions can remain relatively strong. Such variability can create more granular imbalance dynamics, which may be expressed first in the Quarter-Hourly contract. October, at the other extreme, has the lowest endogeneity and the weakest Quarter-Hourly dominance.

Another interesting result is that this seasonal heterogeneity largely disappears under stress. The range of the asymmetry index across months falls from 0.361 in the Normal regime to 0.047 under Stress, and the range of $\Psi_{Q\to H}$ falls from 7.67 to 1.18. The same holds for endogeneity, whose range narrows from 0.023 to 0.007. Under stress the four months look structurally alike, with an asymmetry index between $-0.16$ and $-0.21$ in every case.

The two observations are connected. The size of the regime effect on the asymmetry index reported in Table~\ref{tab:paired} is almost perfectly inversely related to the level of asymmetry in the Normal regime: the correlation across the four months is $-0.99$. The months that are most Quarter-Hourly dominated in calm conditions, April and January, show the largest compression under stress, at $+0.273$ and $+0.219$; the months that are already close to symmetric, July and October, show the smallest, at $+0.083$ and $+0.077$. Stress therefore does not shift the market by a fixed amount. It moves the market towards a common configuration in which the two contracts transmit to each other at comparable strength, irrespective of the seasonal structure that prevails in calm conditions.

This has a direct implication for how the calm-period structure should be used. The leadership of the Quarter-Hourly contract is a feature of normal trading that varies with the season, not a stable property of the market. It is strongest exactly where it is least robust, in the sense that the months in which it is most pronounced are the months in which stress erodes it the most.

\subsection{Response Speed and Shock Persistence}
\label{sec:response}

The branching matrix measures how much order flow one event ultimately generates, but not how quickly. Table~\ref{tab:response} reports, for each transmission channel, the time within which half and ninety per cent of the total excitation is delivered, and the mean excitation lag.

\begin{table*}[htbp]
\centering
\caption{Speed and persistence of excitation by channel and regime. $t_{50}$ and $t_{90}$ are the elapsed times within which 50\% and 90\% of a channel's total excitation has been delivered; the mean lag is the first moment of the normalised kernel. All values are in seconds and are medians across contracts.}
\vspace{10pt}
\label{tab:response}
\small
\begin{tabular}{@{}lccccccc@{}}
\toprule
& \multicolumn{3}{c}{\textbf{Normal}} & \multicolumn{3}{c}{\textbf{Stressed}} & \\
\cmidrule(lr){2-4}\cmidrule(lr){5-7}
\textbf{Channel} & $t_{50}$ & $t_{90}$ & Mean lag &
$t_{50}$ & $t_{90}$ & Mean lag & \textbf{Speed-up} \\ \midrule
$H \rightarrow H$ & 4.15  & 29.0 & 13.26 & 3.61 & 23.5 & 10.36 & $1.3\times$ \\
$H \rightarrow Q$ & 0.098 & 0.59 & 0.24  & 0.17 & 1.37 & 0.50  & $0.5\times$ \\
$Q \rightarrow H$ & 0.68  & 17.8 & 6.05  & 0.26 & 2.38 & 0.93  & $6.5\times$ \\
$Q \rightarrow Q$ & 10.23 & 76.6 & 30.74 & 8.31 & 54.2 & 20.54 & $1.5\times$ \\
\bottomrule
\end{tabular}
\end{table*}

\begin{figure*}[!htbp]
    \centering
    \includegraphics[width=\textwidth]{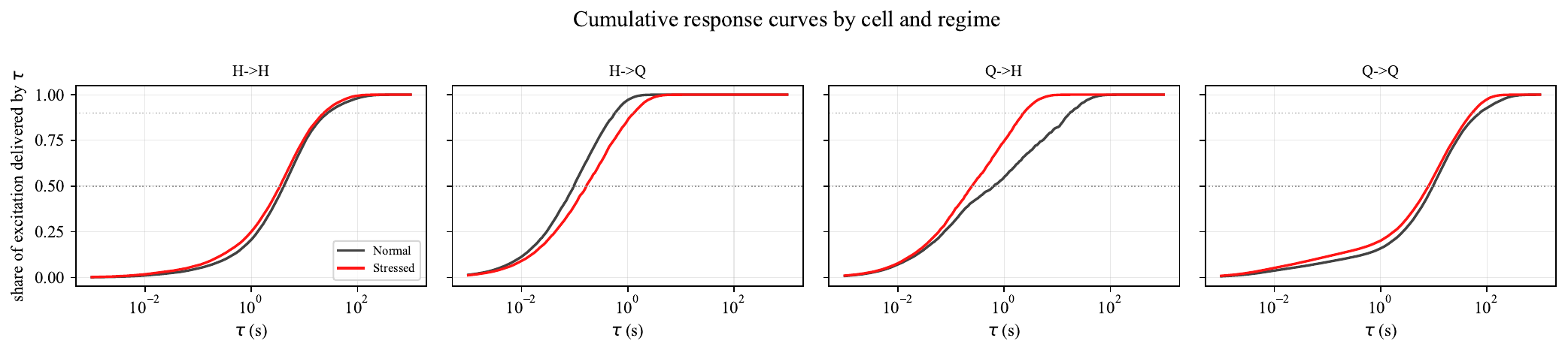}
    \caption{Cumulative response curves. Each panel shows the share of a channel's total excitation delivered within $\tau$ seconds of the triggering event, for the Normal and Stressed regimes. The dotted horizontal lines mark the 50\% and 90\% levels. A curve that rises further to the left indicates faster transmission.}
    \label{fig:response}
\end{figure*}

The four channels operate on clearly separated timescales. Transmission from the Hourly to the Quarter-Hourly contract is by far the fastest: half of the effect is delivered within a tenth of a second and ninety per cent within six tenths of a second. This is the timescale of automated quoting, and it indicates that Quarter-Hourly quotes are updated mechanically in response to Hourly activity. Transmission in the opposite direction is an order of magnitude slower under Normal conditions, with a mean lag of six seconds and ninety per cent of the effect requiring almost eighteen seconds. Self excitation is slower still, particularly within the Quarter-Hourly book, where the mean lag is thirty seconds and 90\% of the excitation is delivered within more than a minute.

The asymmetry in speed complements the asymmetry in magnitude documented above. The Quarter-Hourly contract transmits more to the Hourly contract in total, but does so slowly; the Hourly contract transmits less, but almost instantaneously. The two channels therefore represent different mechanisms. The fast $H \rightarrow Q$ channel is consistent with automated requoting of the granular contract when the principal contract moves. The slow $Q \rightarrow H$ channel is consistent with the gradual absorption of imbalance information by participants who must first observe it accumulate in the Quarter-Hourly book.

Market stress changes this picture substantially, and again the change is concentrated in the $Q \rightarrow H$ channel. Its mean lag collapses from 6.05 to 0.93 seconds, a factor of 6.5, and the time to ninety per cent transmission falls from 17.8 to 2.4 seconds. Under stress, information expressed in the Quarter-Hourly book reaches the Hourly book within seconds rather than tens of seconds. Self-excitation also accelerates, though far less markedly, by factors of 1.3 and 1.5 for the Hourly and Quarter-Hourly books, respectively. The only channel that slows under stress is $H \rightarrow Q$, whose mean lag doubles from 0.24 to 0.50 seconds while its total magnitude increases.

Taken together with the spillover results, this gives a consistent characterisation of the Stressed regime. Market stress does not make the intraday market appreciably more reflexive: the share of endogenous order flow is essentially unchanged at around 95\%. What changes is the structure of transmission between the two contracts. The hierarchy weakens, as the Hourly contract becomes a stronger source of spillover, and the slow channel from the Quarter-Hourly to the Hourly book becomes several times faster. Shocks therefore propagate across contracts both more symmetrically and more quickly under stress, which is the mechanism through which localised liquidity events in one contract can become market-wide events.

\subsubsection{The Composition of Excitation}

\begin{figure*}[!htbp]
    \centering
    \includegraphics[width=\textwidth]{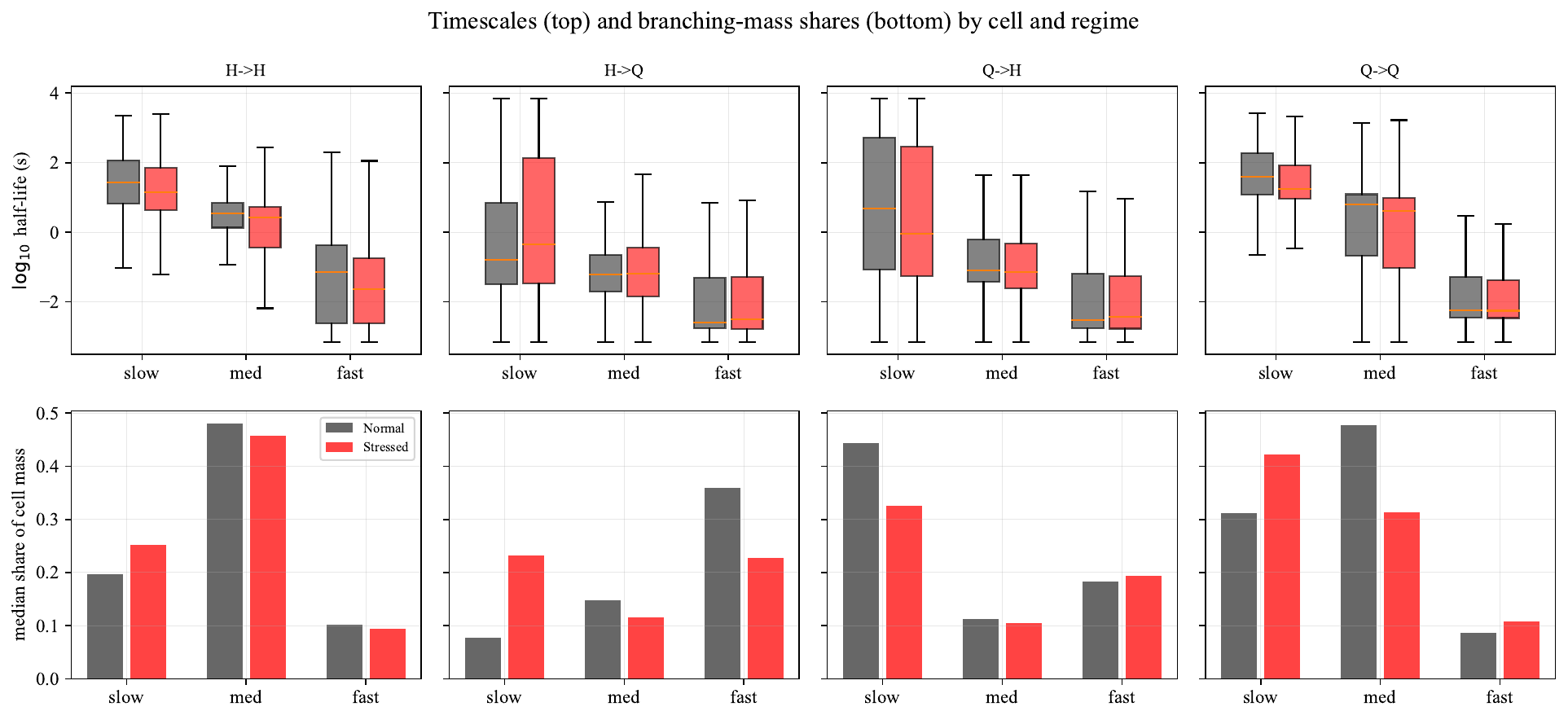}
    \caption{Composition of the estimated kernels. The upper row shows the distribution across contracts of the half-life of each component, on a logarithmic scale; the lower row shows the median share of the channel's total branching mass carried by each component. Slow, medium and fast refer to the three exponential components ordered by decay rate. Shares are medians taken separately for each component and therefore need not sum to one.}
    \label{fig:components}
\end{figure*}

The changes in response speed are not a uniform acceleration of the market. They arise from a reallocation of excitation between the three timescales of the kernel. Figure~\ref{fig:components} shows, for each channel and regime, the estimated half-life of the slow, medium and fast components and the share of the channel's total branching mass that each carries.

The two cross-product channels reallocate mass in opposite directions under stress. In the $H \rightarrow Q$ channel the share carried by the fast component falls from 0.360 to 0.228 while the share carried by the slow component rises from 0.077 to 0.232; in the $Q \rightarrow H$ channel the slow share falls from 0.444 to 0.325. The half-lives move the same way: the slow component of the $Q \rightarrow H$ channel accelerates from 4.85 to 0.92 seconds, a factor of 5.2, while the slow component of the $H \rightarrow Q$ channel decelerates from 0.158 to 0.459 seconds. Stress therefore does not simply speed the market up. It moves transmission from the Quarter-Hourly to the Hourly contract out of a slow channel and into faster ones, and moves transmission in the opposite direction partly out of its fast channel into a slower one. This is the mechanism underlying both the compression of the asymmetry index and the pattern of speed changes in Table~\ref{tab:response}.

Within each book the reallocation is milder and in the same direction for both contracts: the slow share rises under stress, from 0.196 to 0.252 in the $H \rightarrow H$ channel and from 0.312 to 0.423 in the $Q \rightarrow Q$ channel, while the half-lives of all components shorten. Self-excitation thus becomes faster in every component while placing somewhat more of its weight on the slowest one, whereas the cross-product channels are restructured.

\subsubsection{Robustness of the Kernel Decomposition}

The three components of each kernel are not equally well identified. The decomposition into slow, medium and fast timescales is stable for the self-excitation channels, where fewer than 5\% of contracts have a decay rate resting on a bound. For the cross-product channels the slow component is frequently inactive: its magnitude $\alpha$ is estimated at the lower bound for 48.3\% of contracts in the $H \rightarrow Q$ channel and 49.4\% in the $Q \rightarrow H$ channel under Normal conditions. In these contracts the model attributes no slow cross-excitation at all, and cross-product transmission operates entirely through the medium and fast components. The fast component reaches the upper bound on the decay rate in 12 to 15\% of contracts in the cross-product channels, so its timescale is not resolved beyond the granularity of the data; the branching mass it carries remains identified and is the quantity reported above. The upper row of Figure~\ref{fig:components} shows the consequence directly: the interquartile range of the slow component's half-life spans 2.4 decades in the $H \rightarrow Q$ channel and 3.8 decades in the $Q \rightarrow H$ channel, against approximately one decade for the corresponding component of the $Q \rightarrow Q$ channel. These features do not affect the aggregate quantities of Tables~\ref{tab:branching}--\ref{tab:response}, which depend on the integrated kernel rather than on the allocation of mass across components, but they do mean that the individual component timescales in the cross-product channels should be read as indicative rather than precise.

\section{Conclusion and Discussion}
\label{sec:conclusion}

This paper studies how liquidity shocks travel between contracts of different granularity in the IDC market. Trading in the same delivery period takes place simultaneously in Hourly and Quarter-Hourly products, and activity in one book can trigger activity in the other. We ask how strong this cross-product transmission is, in which direction it runs, how long it takes, and whether it changes when liquidity deteriorates.

We model the two order books jointly as a SDHP. The baseline intensity and the excitation kernel both depend on an observed liquidity regime, so that the exogenous arrival of orders and the market's reaction to its own past are estimated separately in calm and stressed conditions. We compare four excitation kernels within this framework, three exponential mixtures and one power law, and derive the analytical gradients of the multi-session log-likelihood and of the stability penalty for both kernel families, which is what makes estimation at the scale of individual contracts possible. The model is estimated separately for each of 2,915 delivery contracts in the first month of each quarter of 2024, so that regimes are compared within the same contract and the results can be examined across delivery hours, days and seasons. The three-component exponential kernel performs best overall across the criteria and is used throughout.

The market turns out to be strongly self-driven. About ninety-five per cent of order flow is a reaction to earlier order flow, and only about five per cent arrives from outside the book, in both regimes. Because each book amplifies its own activity so heavily, a weak direct link between the two contracts produces a large cumulative effect: a first-round cross-excitation of about 0.02 orders corresponds to more than two orders once the full chain of reactions is counted. Cross-product effects that look negligible in the raw coefficients are in fact economically important.

Transmission between the two contracts is uneven. In calm conditions the Quarter-Hourly contract leads: one order there eventually generates 6.68 Hourly orders, against 2.39 in the opposite direction. This fits the role of the granular contract as the place where imbalance appears first. Under stress the gap narrows from 2.8 to 1.5, and the slow channel running from the Quarter-Hourly to the Hourly book speeds up, with its mean lag falling from six seconds to under one. Stress does not make the market more self-driven; the endogenous share barely moves. What changes is how the two books are connected. They become more equal and react to each other faster, which is how a liquidity problem in one contract can spread to the other.

This leadership is also seasonal. It is strongest in April and January and much weaker in July and October, but under stress all four months converge to nearly the same configuration, and the months with the strongest calm-period leadership are those where stress erodes it most. The order in which the two contracts respond is therefore a feature of normal trading rather than a fixed property of the market. A participant hedging across the two products cannot rely on it, and it is least dependable during stressed periods.

\subsection{Limitations}
\label{sec:conclusion_limits}

EPEX SPOT records order messages to the millisecond, which is a very limited precision in comparison with the microsecond and nanosecond timestamps available in other markets. A Hawkes process requires a strictly positive time between successive events, so orders arriving less than a millisecond apart cannot be separated. We take the recorded sequence as the true order of arrival, but the spacing between such orders is not observed. The effect is visible in the estimates: for twelve to fifteen per cent of contracts the fastest kernel component rests on the upper bound of the decay rate, because the data cannot resolve excitation faster than the recording granularity. The mass carried by that component is still identified and is what we report, but its timescale is not.

The same limitation is why we model order submissions only. Submissions and cancellations often carry the same timestamp, since cancel-and-replace activity comes from the same algorithm within the same millisecond, so including both would create many simultaneous events. Estimating two separate models would avoid this but would remove the cross-effects between adding and cancelling liquidity, which is the reason for including cancellations in the first place. Finer timestamps, or a model that admits simultaneous events, would be needed.

The power-law kernel is estimated on a truncated excitation history, whereas the exponential kernels are exact. This is partly a property of the kernel: as shown in Section~\ref{sec:compcost}, the power law has no recursive representation, so its intensity depends on the whole history and exact evaluation is quadratic in the number of events. It is also partly a practical matter, since we estimate it with analytical gradients and a longer history is affordable at some cost in computation. The truncation means the power-law likelihood is not directly comparable with the exponential ones, which is why the kernel comparison rests on residual diagnostics instead. Since the power law was not selected, this does not affect the results, but it does limit its use as a robustness check.

Finally, we cover only the Hourly and Quarter-Hourly products. The market also lists Half-Hourly contracts, which we exclude because they are much less liquid and contract-level estimation needs enough events in each regime.

\section{Acknowledgments}
This research is fully funded by the Starter Grant awarded to Dr. Tarek Alskaif and Dr. Kwabena Bennin at the Information Technology Group, Wageningen University. It is partly supported by Project HighLO, a collaboration between Wageningen University \& Research (WUR) and the European Organization for Nuclear Research (CERN), which aims to identify malicious trading behavior in electricity markets. During the preparation of this work, the authors used Claude (Anthropic) in order to assist with text editing, language refinement, and improving the clarity and structure of the manuscript. After using this tool, the authors reviewed and edited the content as needed and take full responsibility for the content of the published article.

% NOTE this caused a compile warning
\bibliographystyle{IEEEtran}
\bibliography{references}
\onecolumn
\appendix
\section{Goodness of Fit via Time-Rescaled Residuals}
\label{app:time_rescaling}
 
This appendix details the goodness-of-fit procedure used to evaluate the estimated point-process models. The procedure rests on the random time change theorem, which converts any simple point process with a strictly positive conditional intensity into a unit-rate Poisson process. Two features make it well suited to the present comparison. First, it applies to \emph{any} conditional intensity, so it is common to all four kernel specifications and requires no modification from one
to the next. Second, the resulting null distribution is free of the model parameters, so the diagnostic does not depend on the likelihood being evaluated exactly---a property that matters here, since the power-law likelihood is computed on a truncated kernel while the exponential likelihoods are exact.
 
% ------------------------------------------------------------
\subsection{Random time change}
 
Let $N(t)$ be a simple point process adapted to a history $\mathcal{F}_t$, with bounded, strictly positive conditional $\mathcal{F}$-intensity $\lambda^*(t)$ and compensator
\begin{equation}
\Lambda^*(t) = \int_0^t \lambda^*(u)\,du,
\label{eq:app_compensator}
\end{equation}
which is not almost surely bounded. The compensator is the cumulative expected number of events up to time $t$, conditional on the past.
 
For a homogeneous Poisson process of unit rate, interarrival times are independent and exponentially distributed with unit mean. For an inhomogeneous or self-exciting process the intensity fluctuates, but a random change of time scale restores this structure. Intuitively, measuring time in units of $\Lambda^*$ rather than $t$ stretches the clock during periods of high intensity and compresses it when the intensity is low: clustered events are pulled apart, sparse events drawn together.
 
\begin{theorem}[Random time change]
Under the conditions above, the time-changed process
\begin{equation}
\tilde{N}(t) = N\big(\Lambda^{*-1}(t)\big)
\label{eq:rtc_appendix}
\end{equation}
is a Poisson process of unit rate. Equivalently, the transformed interarrival times are independent and identically distributed $\mathrm{Exp}(1)$ random variables.
\end{theorem}
 
\begin{remark}[Multivariate case]
The result extends to multivariate point processes: if each component $N_e$ is rescaled by its \emph{own} compensator $\Lambda^*_e$, the resulting processes are \emph{independent} unit-rate Poisson processes. This is what licenses pooling residuals across event types, and across regimes and sessions, when assessing the $\mathrm{Exp}(1)$ null.
\end{remark}
 
\subsection{Residual construction}
 
Fix a session $j$ and an event type $e$, and let $t^{(j)}_{1,e}<t^{(j)}_{2,e}<\dots$ denote the observed times of type-$e$ events in that session. The time-rescaled residuals are the increments of the compensator between consecutive events,
\begin{equation}
r^{(j)}_{n,e}
=
\int_{t^{(j)}_{n-1,e}}^{t^{(j)}_{n,e}}
\lambda^{(j)}_e(t)\,dt ,
\label{eq:app_residual}
\end{equation}
where $\lambda^{(j)}_e$ is the fitted conditional intensity \eqref{eq:sdhp_intensity} evaluated with the estimated parameters $\hat\theta$. The lower limit for the first residual is the end of the burn-in window rather than the session open, so that the intensity has accumulated a representative history before any residual is formed. The interval following the last observed event is right-censored by the session close $T_j$, is therefore not $\mathrm{Exp}(1)$, and is discarded.
Under correct specification the $r^{(j)}_{n,e}$ are i.i.d.\ $\mathrm{Exp}(1)$; equivalently $u^{(j)}_{n,e}=1-\exp(-r^{(j)}_{n,e})$ are i.i.d.\ uniform on $[0,1]$.
 
Because the model is regime-dependent, we additionally form regime-conditional residuals. For a fixed regime $x$, the type-$e$ events occurring while $S^{(j)}(t^-)=x$ constitute a point process whose conditional intensity is $\mathbf{1}\{S^{(j)}(t^-)=x\}\,\lambda^{(j)}_e(t)$; this is again predictable, since the regime path is observed and evaluated at $t^-$. Let $t^{(j,x)}_{1,e}<t^{(j,x)}_{2,e}<\dots$ denote the times of successive type-$e$ events that occur while the market is in regime $x$. The corresponding residuals are
\begin{equation}
r^{(j,x)}_{k,e}
=
\int_{t^{(j,x)}_{k-1,e}}^{t^{(j,x)}_{k,e}}
\mathbf{1}\{S^{(j)}(t^-)=x\}\,\lambda^{(j)}_e(t)\,dt .
\label{eq:app_residual_state}
\end{equation}
The integral spans the entire interval between two consecutive regime-$x$ events of type $e$, including any intervening excursions into the other regime, which contribute nothing on account of the indicator. By the same theorem, these residuals are i.i.d.\ $\mathrm{Exp}(1)$ under correct regime-wise specification.

\subsection{Statistical evaluation}
 
Two distinct implications of the null are assessed: that the residuals are marginally $\mathrm{Exp}(1)$, and that they are serially independent. A model may satisfy one and violate the other, so both are reported.
 
The marginal distribution is assessed graphically by quantile--quantile plots against the $\mathrm{Exp}(1)$ law, and numerically by the Kolmogorov--Smirnov distance between the empirical distribution of the residuals and the $\mathrm{Exp} (1)$ distribution function, together with their sample mean and variance, both equal to unity under the null. Serial independence is assessed by the lag-one autocorrelation of the residuals, which is zero under the null and which the marginal diagnostics cannot detect.
 
We report these quantities as \emph{discrepancy measures} rather than as test statistics, and we do not attach $p$-values to them. Two considerations motivate this choice. First, the number of residuals is of order $10^{6}$, at which sample size any formal test of distributional fit rejects for arbitrarily small departures from the null; a rejection would therefore convey no information about the relative adequacy of the four specifications. Second, the residuals are constructed from an intensity evaluated at estimated rather than known parameters, so the null distributions of the Kolmogorov--Smirnov and Anderson--Darling statistics are not those tabulated for a fully specified hypothesis. Comparing the magnitudes of these statistics across specifications remains informative, whereas comparing them against critical values does not.
 
Finally, since pooled residuals can conceal heterogeneity across trading sessions, we also report the distribution of the per-session Kolmogorov--Smirnov distance for each specification, which reveals whether a model that fits well on average deteriorates systematically on the stressed sessions of primary interest.

\section{Analytical Gradients}
\label{app:gradients}
 
This appendix derives the gradients of the multi-session log-likelihood $\mathcal{L}$ and of the per-state spectral-radius penalty $\mathcal{P}$ with respect to the parameters of both kernel families. Throughout, $j=1,\dots,J$ indexes sessions, $r=1,\dots,R$ indexes exponential components, and $S^{(j)}(t)$ denotes the regime prevailing at time $t$ in session $j$, evaluated at $t^-$.
 
Because the conditional intensity \eqref{eq:sdhp_intensity} is state-filtered, differentiation with respect to a parameter carrying the regime index $x$ produces the indicator $\mathbf{1}\{S^{(j)}(t)=x\}$. This factor appears in \emph{both} the event sums and the compensator integrals below: an event contributes to the gradient of a state-$x$ parameter only if it occurs while the market is in regime $x$.
 
It is convenient to define the exposure time of regime $x$ in session $j$,
\begin{equation}
\mathcal{T}^{(j)}_x=\int_0^{T_j}\mathbf{1}\{S^{(j)}(t)=x\}\,dt,
\qquad x\in\{0,1\}.
\label{eq:app_exposure}
\end{equation}
 
\subsection{Baseline Intensity}
 
Since $\partial\lambda^{(j)}_{e}(t)/\partial\mu_{x,e}=\mathbf{1}\{S^{(j)}(t)=x\}$, the baseline gradient is common to both kernel families:
\begin{equation}
\frac{\partial\mathcal{L}}{\partial \mu_{x,e}}
=
\sum_{j=1}^{J}\sum_{k:\,e^{(j)}_k=e}
\frac{\mathbf{1}\{S^{(j)}(t^{(j)-}_k)=x\}}{\lambda^{(j)}_{e}(t^{(j)}_k)}
\;-\;
\sum_{j=1}^{J}\mathcal{T}^{(j)}_x .
\label{eq:app_dL_dmu}
\end{equation}

\subsection{Exponential-mixture Kernel}
 
\paragraph{History transforms.}
For each $(x,e'\!\to\!e,r)$ define the state-parameterised history sums
\begin{align}
\mathcal{R}^{(x,r),(j)}_{e'\to e}(t)
&=\!\!\sum_{\substack{t^{(j)}_i<t\\ e^{(j)}_i=e'}}
    \exp\!\big(-\beta^{(x,r)}_{e'\to e}\,(t-t^{(j)}_i)\big),
\label{eq:app_R}\\[2pt]
\widetilde{\mathcal{R}}^{(x,r),(j)}_{e'\to e}(t)
&=\!\!\sum_{\substack{t^{(j)}_i<t\\ e^{(j)}_i=e'}}
    (t-t^{(j)}_i)\,\exp\!\big(-\beta^{(x,r)}_{e'\to e}\,(t-t^{(j)}_i)\big).
\label{eq:app_Rtilde}
\end{align}
These relate to the partial sums $\Lambda^{(x,r)}_{e'\to e}$ of the main text by $\Lambda^{(x,r)}_{e'\to e}=\alpha^{(x,r)}_{e'\to e}\,\mathcal{R}^{(x,r)}_{e'\to e}$, and both obey the recursion \eqref{eq:recursion}, so each is evaluated in a single pass through the data.
 
\paragraph{Gradients.}
Since $\partial\lambda^{(j)}_e(t)/\partial\alpha^{(x,r)}_{e'\to e}
=\mathbf{1}\{S^{(j)}(t)=x\}\,\mathcal{R}^{(x,r),(j)}_{e'\to e}(t)$ and
$\partial\lambda^{(j)}_e(t)/\partial\beta^{(x,r)}_{e'\to e}
=-\mathbf{1}\{S^{(j)}(t)=x\}\,\alpha^{(x,r)}_{e'\to e}\,\widetilde{\mathcal{R}}^{(x,r),(j)}_{e'\to e}(t)$,
we obtain

\begin{align}
\frac{\partial \mathcal{L}}{\partial\alpha^{(x,r)}_{e'\to e}}
&=
\sum_{j=1}^{J}\sum_{k:\,e^{(j)}_k=e}
\mathbf{1}\{S^{(j)}(t^{(j)-}_k)=x\}\,
\frac{\mathcal{R}^{(x,r),(j)}_{e'\to e}(t^{(j)}_k)}{\lambda^{(j)}_{e}(t^{(j)}_k)}
\ -\sum_{j=1}^{J}\int_0^{T_j}\mathbf{1}\{S^{(j)}(t)=x\}\,
\mathcal{R}^{(x,r),(j)}_{e'\to e}(t)\,dt,
\label{eq:app_dL_dalpha}
\\[2pt]
\frac{\partial \mathcal{L}}{\partial\beta^{(x,r)}_{e'\to e}}
&=
-\sum_{j=1}^{J}\sum_{k:\,e^{(j)}_k=e}
\mathbf{1}\{S^{(j)}(t^{(j)-}_k)=x\}\,
\frac{\alpha^{(x,r)}_{e'\to e}\,\widetilde{\mathcal{R}}^{(x,r),(j)}_{e'\to e}(t^{(j)}_k)}
     {\lambda^{(j)}_{e}(t^{(j)}_k)}
 +\sum_{j=1}^{J}\int_0^{T_j}\mathbf{1}\{S^{(j)}(t)=x\}\,
\alpha^{(x,r)}_{e'\to e}\,\widetilde{\mathcal{R}}^{(x,r),(j)}_{e'\to e}(t)\,dt,
\label{eq:app_dL_dbeta}
\end{align}
for $r=1,\dots,R$ and $x\in\{0,1\}$.

\subsection{Power-law Kernel}
\paragraph{History transforms.}
Write $\Delta_i = t-t^{(j)}_i$ and, for each $(x,e'\!\to\!e)$, define
\begin{align}
\mathcal{A}^{(x),(j)}_{e'\to e}(t)
&=\!\!\sum_{\substack{t^{(j)}_i<t\\ e^{(j)}_i=e'}}
   \big(\Delta_i+c^{(x)}_{e'\to e}\big)^{-\eta^{(x)}_{e'\to e}},
\label{eq:app_A}\\
\mathcal{B}^{(x),(j)}_{e'\to e}(t)
&=\!\!\sum_{\substack{t^{(j)}_i<t\\ e^{(j)}_i=e'}}
   \big(\Delta_i+c^{(x)}_{e'\to e}\big)^{-\eta^{(x)}_{e'\to e}}
   \log\!\big(\Delta_i+c^{(x)}_{e'\to e}\big),
\label{eq:app_B}\\
\mathcal{C}^{(x),(j)}_{e'\to e}(t)
&=\!\!\sum_{\substack{t^{(j)}_i<t\\ e^{(j)}_i=e'}}
   \big(\Delta_i+c^{(x)}_{e'\to e}\big)^{-\eta^{(x)}_{e'\to e}-1}.
\label{eq:app_C}
\end{align}
In practice the sums in \eqref{eq:app_A}--\eqref{eq:app_C} are restricted to $\Delta_i\le\Delta_{\max}$, consistent with the truncation applied to the intensity itself. Note that $\mathcal{A}$, $\mathcal{B}$ and $\mathcal{C}$ share the factor
$(\Delta_i+c)^{-\eta}$, so all three are accumulated in one pass.
 
\paragraph{Gradients.}
Differentiating $\kappa^{(x)}_{e'\to e}(\tau)=\alpha^{(x)}_{e'\to e}(\tau+c^{(x)}_{e'\to e})^{-\eta^{(x)}_{e'\to e}}$
gives
$\partial\kappa/\partial\alpha=(\tau+c)^{-\eta}$,
$\partial\kappa/\partial\eta=-\alpha(\tau+c)^{-\eta}\log(\tau+c)$ and
$\partial\kappa/\partial c=-\alpha\,\eta\,(\tau+c)^{-\eta-1}$, whence \eqref{eq:app_pow_dalpha}--\eqref{eq:app_pow_dc}

\begin{align}
\frac{\partial \mathcal{L}}{\partial\alpha^{(x)}_{e'\to e}}
&=
\sum_{j=1}^{J}\sum_{k:\,e^{(j)}_k=e}
\mathbf{1}\{S^{(j)}(t^{(j)-}_k)=x\}\,
\frac{\mathcal{A}^{(x),(j)}_{e'\to e}(t^{(j)}_k)}{\lambda^{(j)}_{e}(t^{(j)}_k)}
\;-\;\sum_{j=1}^{J}\int_0^{T_j}\!\!\mathbf{1}\{S^{(j)}(t)=x\}\,
\mathcal{A}^{(x),(j)}_{e'\to e}(t)\,dt,
\label{eq:app_pow_dalpha}
\\[4pt]
\frac{\partial \mathcal{L}}{\partial\eta^{(x)}_{e'\to e}}
&=
-\alpha^{(x)}_{e'\to e}\!
\left[
\sum_{j=1}^{J}\sum_{k:\,e^{(j)}_k=e}
\mathbf{1}\{S^{(j)}(t^{(j)-}_k)=x\}\,
\frac{\mathcal{B}^{(x),(j)}_{e'\to e}(t^{(j)}_k)}{\lambda^{(j)}_{e}(t^{(j)}_k)}
-\sum_{j=1}^{J}\int_0^{T_j}\!\!\mathbf{1}\{S^{(j)}(t)=x\}\,
\mathcal{B}^{(x),(j)}_{e'\to e}(t)\,dt
\right],
\label{eq:app_pow_deta}
\\[4pt]
\frac{\partial \mathcal{L}}{\partial c^{(x)}_{e'\to e}}
&=
-\alpha^{(x)}_{e'\to e}\,\eta^{(x)}_{e'\to e}\!
\left[
\sum_{j=1}^{J}\sum_{k:\,e^{(j)}_k=e}
\mathbf{1}\{S^{(j)}(t^{(j)-}_k)=x\}\,
\frac{\mathcal{C}^{(x),(j)}_{e'\to e}(t^{(j)}_k)}{\lambda^{(j)}_{e}(t^{(j)}_k)}
-\sum_{j=1}^{J}\int_0^{T_j}\!\!\mathbf{1}\{S^{(j)}(t)=x\}\,
\mathcal{C}^{(x),(j)}_{e'\to e}(t)\,dt
\right].
\label{eq:app_pow_dc}
\end{align}

\subsection{Penalty Gradient}
\label{app:penalty_gradients}
 
The penalty depends on $\theta$ only through the branching matrices $G^{(0)},G^{(1)}$, so its gradient factorises into three stages: the derivative of $\mathcal{P}$ with respect to $\rho_x$, the derivative of $\rho_x$ with respect to $G^{(x)}$, and the derivative of $G^{(x)}$ with respect to the kernel parameters. Only the last stage is kernel-specific.
 
\paragraph{Stage 1:}
Writing $\Delta_x:=\rho_x-\rho^\star$ and $s_x:=\max\{0,\Delta_x\}$, the penalty
\eqref{eq:penalty-2states} is $\mathcal{P}(\theta)=\sum_{x}\omega_x s_x^2$ and
\begin{equation}
\frac{\partial \mathcal{P}}{\partial \rho_x}
=
2\omega_x\, s_x\, \mathbf{1}\{\Delta_x>0\}.
\label{eq:app_dP_drho}
\end{equation}
The penalty is therefore inactive, with zero gradient, whenever $\rho_x\le\rho^\star$.
 
\paragraph{Stage 2:}
Let $\rho_x$ be a simple eigenvalue of $G^{(x)}$, with right and left eigenvectors $v^{(x)}$ and
$u^{(x)}$,
\[
G^{(x)} v^{(x)} = \rho_x v^{(x)},
\qquad
\big(u^{(x)}\big)^\top G^{(x)} = \rho_x \big(u^{(x)}\big)^\top,
\]
normalised so that $\big(u^{(x)}\big)^\top v^{(x)} = 1$. The entrywise derivative is then
\begin{equation}
\frac{\partial \rho_x}{\partial G^{(x)}_{e'e}} = u^{(x)}_{e'}\, v^{(x)}_{e},
\label{eq:app_drho_dG}
\end{equation}
and, by the chain rule,
\begin{equation}
\frac{\partial \mathcal{P}}{\partial G^{(x)}_{e'e}}
=
2\omega_x\, s_x\, \mathbf{1}\{\Delta_x>0\}\; u^{(x)}_{e'} v^{(x)}_{e}.
\label{eq:app_dP_dG}
\end{equation}
Because the estimated kernels have strictly positive amplitudes, $G^{(x)}$ has strictly positive entries; by the Perron--Frobenius theorem its spectral radius is then a simple, positive eigenvalue with positive left and right eigenvectors, so \eqref{eq:app_drho_dG} is well defined and $u^{(x)}_{e'}v^{(x)}_{e}>0$.
 
\paragraph{Stage 3a: exponential-mixture kernel.}
From $G^{(x)}_{e'e}=\sum_{r=1}^{R}\alpha^{(x,r)}_{e'\to e}\big/\beta^{(x,r)}_{e'\to e}$,
\begin{equation}
\frac{\partial G^{(x)}_{e'e}}{\partial \alpha^{(x,r)}_{e'\to e}}
=
\frac{1}{\beta^{(x,r)}_{e'\to e}},
\qquad
\frac{\partial G^{(x)}_{e'e}}{\partial \beta^{(x,r)}_{e'\to e}}
=
-\frac{\alpha^{(x,r)}_{e'\to e}}{\big(\beta^{(x,r)}_{e'\to e}\big)^2},
\label{eq:app_dG_dab}
\end{equation}
and combining with \eqref{eq:app_dP_dG},
\begin{align}
\frac{\partial \mathcal{P}}{\partial \alpha^{(x,r)}_{e'\to e}}
&=
2\omega_x\, s_x\, \mathbf{1}\{\Delta_x>0\}\;
u^{(x)}_{e'} v^{(x)}_{e}\;
\frac{1}{\beta^{(x,r)}_{e'\to e}},
\label{eq:app_dP_dalpha}
\\[4pt]
\frac{\partial \mathcal{P}}{\partial \beta^{(x,r)}_{e'\to e}}
&=
-2\omega_x\, s_x\, \mathbf{1}\{\Delta_x>0\}\;
u^{(x)}_{e'} v^{(x)}_{e}\;
\frac{\alpha^{(x,r)}_{e'\to e}}{\big(\beta^{(x,r)}_{e'\to e}\big)^2}.
\label{eq:app_dP_dbeta}
\end{align}
 
\paragraph{Stage 3b: power-law kernel.}
From 
\\ $G^{(x)}_{e'e}=\alpha^{(x)}_{e'\to e}\big(c^{(x)}_{e'\to e}\big)^{1-\eta^{(x)}_{e'\to e}}
\big/\big(\eta^{(x)}_{e'\to e}-1\big)$, suppressing indices for readability,
\begin{equation}
\begin{gathered}
\frac{\partial G}{\partial \alpha}=\frac{G}{\alpha}, \\
\frac{\partial G}{\partial \eta}=-G\left(\log c+\frac{1}{\eta-1}\right), \\
\frac{\partial G}{\partial c}=-\,\frac{G\,(\eta-1)}{c}.
\end{gathered}
\label{eq:app_dG_pow}
\end{equation}
each valid for $\eta>1$, and the penalty gradients follow from \eqref{eq:app_dP_dG} exactly as
above. Expression \eqref{eq:app_dG_pow} makes the singularity as $\eta\downarrow 1$ explicit: both
$G$ and its derivative with respect to $\eta$ diverge, which motivates bounding $\log(\eta-1)$ away
from $-\infty$ during optimisation.
 
\paragraph{Baseline parameters.}
Since $\mathcal{P}$ depends on $\theta$ only through $G^{(x)}$,
\[
\frac{\partial \mathcal{P}}{\partial \mu_{x,e}} = 0
\qquad \text{for all } x\in\{0,1\},\ e\in\mathcal{E}.
\]
 
\subsection{Reparametrised Gradients for The Power-law Kernel}
 
The power-law model is estimated in the coordinates
$\big(\log\mu,\ \log\alpha,\ \log(\eta-1),\ \log c\big)$. Gradients transform by the chain rule,
\begin{align}
\frac{\partial \mathcal{J}}{\partial \log\mu}
&=
\mu\,\frac{\partial \mathcal{J}}{\partial \mu},
\qquad
\frac{\partial \mathcal{J}}{\partial \log\alpha}
=
\alpha\,\frac{\partial \mathcal{J}}{\partial \alpha},
\\
\frac{\partial \mathcal{J}}{\partial \log(\eta-1)}
&=
(\eta-1)\,\frac{\partial \mathcal{J}}{\partial \eta},
\qquad
\frac{\partial \mathcal{J}}{\partial \log c}
=
c\,\frac{\partial \mathcal{J}}{\partial c},
\label{eq:app_logspace}
\end{align}
applied entrywise to $\mathcal{J}=\mathcal{L}-\mathcal{P}$.

\section{Initialization and Optimization}
\label{app:estimation}
The log-likelihood surface of a high-dimensional, multi-exponential Hawkes
process is highly non-convex, so the result of an optimisation depends on where it starts. Rather than impose an ordering of timescales through a fixed
starting point, we draw the starting values at random within economically
admissible bounds, fit each contract independently, and run the optimiser to
convergence. This keeps the estimated decomposition of timescales a property of
the data rather than of the initialisation.

To capture the full spectrum of market memory, from macro-level human trading to millisecond bursts, we bound the decay parameters over a wide admissible range. We set $\beta_{\min} = 10^{-4}$ and $\beta_{\max} = 1000.0$, where the upper bound accommodates reactions on the order of $1/\beta_{\max} = 1$ millisecond, which is the lowest resolution of EPEX SPOT data. For a kernel with $R$ exponentials, we initialize the decay rates by drawing them at random, log-uniformly over this range, so that the starting point does not impose an ordering of timescales. Let $r \in \{1,\dots,R\}$ index the exponentials; for each cell the draws are sorted in increasing order, so that $r$ indexes the kernels from slow to fast:
\begin{equation}
\log \beta^{(r),(0)}_{x,e'\rightarrow e} \sim \mathcal{U}\big(\log \beta_{\min},\, \log \beta_{\mathrm{top}}\big), \qquad x \in \{0,1\},
\label{eq:beta-init-triexp}
\end{equation}
where the sampling ceiling $\beta_{\mathrm{top}}$ is taken from the smallest observed inter-event time rather than from $\beta_{\max}$. A component initialized at a rate so high that $e^{-\beta\tau}$ underflows for every observed lag lies in a region where the likelihood is flat in both $\alpha$ and $\beta$, and no gradient-based method can move it; keeping $\beta_{\mathrm{top}}$ inside the resolvable range ensures that every kernel starts where the data can inform it, while $\beta_{\max}$ continues to act as a guard.

The baseline intensities are seeded from the unconditional empirical event rates. For each event type $e$, let $\nu_e$ be the unconditional rate pooled over the fitted contract. To break symmetry and reflect typical regime behaviour, we initialize the Normal state (State~0) baselines at $0.5\,\nu_e$ and the Stressed state (State~1) baselines at $1.5\,\nu_e$:
\begin{equation}
\mu^{(0)}_{0,e} = 0.5\,\nu_e, \qquad \mu^{(0)}_{1,e} = 1.5\,\nu_e.
\end{equation}
To keep the initial branching ratio safely subcritical and stable, the excitation coefficients ($\alpha$) are initialized from a randomly drawn branching mass. For each cell and kernel we draw a target branching ratio $b^{(p),(0)}_{x,e'\rightarrow e} \sim \mathcal{U}(0.01,\,0.30)$, divide it evenly across the $R$ exponentials, and scale by the corresponding decay rate:
\begin{equation}
\alpha^{(r),(0)}_{x,e'\rightarrow e} = \frac{b^{(r),(0)}_{x,e'\rightarrow e}}{P}\, \beta^{(r),(0)}_{x,e'\rightarrow e}, \qquad r\in\{1,\dots,R\},\ x\in\{0,1\}.
\label{eq:alpha-init-triexp}
\end{equation}
Drawing the masses independently across cells means that a kernel which fails to move from its starting point does not contribute the same value to every entry of the branching matrix, which makes such a component detectable by comparison with its initial value. We verify for each reported fit that the kernels have moved away from their initialization.

Minimization of the full penalized objective $\mathcal{J}(\theta)$ from \eqref{eq:objective} is carried out with the L-BFGS-B quasi-Newton algorithm in the logarithmic coordinates of Section~\ref{subsec:estimation}, from a single random initialization run to convergence. To accommodate the extreme clustering of the hourly contracts without log-likelihood underflow or spectral explosion, we enforce the following box constraints during optimization:
$\mu_{x,e} \in [10^{-4},\, \max(100\,\mu^{(0)}_{x,e},\, 10.0)]$,
$\alpha^{(r)}_{x,e'\rightarrow e} \in [10^{-4},\, 500.0]$,
and $\beta^{(r)}_{x,e'\rightarrow e} \in [10^{-4},\, 1000.0]$,
for all states $x\in\{0,1\}$ and exponential kernels $r\in\{1,\dots,R\}$.

\end{document}